\documentclass[twocolumn,tighten,astrosymb]{aastex631}

\usepackage[version=4]{mhchem}

\newcommand{\Kdeg}{\,\text{K}}

\newcommand{\au}{\,\text{au}}
\newcommand{\yr}{\,\text{yr}}

\newcommand{\kms}{\,\text{km\,s}^{-1}}

\newcommand{\cmcms}{\,\text{cm}^{2}\,\text{s}^{-1}}
\newcommand{\Msun}{\,M_{\odot}}

\newcommand{\erg}{\,\text{erg}}

\graphicspath{{./}{figure/}}

\shortauthors{omura et al.}
\shorttitle{outflow and jet from HH270mms1}

\accepted{December 29, 2023}

\begin{document}
\title{Revealing multiple nested molecular outflows with rotating signatures in HH270mms1-A with ALMA}

\correspondingauthor{Mitsuki Omura}
\email{omura.mitsuki.362@s.kyushu-u.ac.jp}

\author[0000-0002-7951-1641]{Mitsuki Omura}
\affiliation{Department of Earth and Planetary Sciences, Graduate School of Science, Kyushu University, 744 Motooka, Nishi-ku, Fukuoka 819-0395, Japan}
\author[0000-0002-2062-1600]{Kazuki Tokuda}
\affiliation{Department of Earth and Planetary Sciences, Faculty of Science, Kyushu University, 744 Motooka, Nishi-ku, Fukuoka 819-0395, Japan}
\affiliation{National Astronomical Observatory of Japan, National Institutes of Natural Sciences, 2-21-1 Osawa, Mitaka, Tokyo 181-8588, Japan}
\author[0000-0002-0963-0872]{Masahiro N. Machida}
\affiliation{Department of Earth and Planetary Sciences, Faculty of Science, Kyushu University, 744 Motooka, Nishi-ku, Fukuoka 819-0395, Japan}

\begin{abstract}

We present molecular line observations of the protostellar outflow associated with HH270mms1 in the Orion B molecular cloud with ALMA. 
The \ce{^12CO}($J$=\,3\,--\,2) emissions show that the outflow velocity structure consists of four distinct components of low ($\lesssim 10\kms$), intermediate ($\sim10-25\kms$) and high ($\gtrsim40\kms$) velocities in addition to the entrained gas velocity ($\sim25-40\kms$).
The high- and intermediate-velocity flows have well-collimated structures surrounded by the low-velocity flow. 
The chain of knots is embedded in the high-velocity flow or jet, which is the evidence of episodic mass ejections induced by time-variable mass accretion. 
We could detect the velocity gradients perpendicular to the outflow axis in both the low- and intermediate-velocity flows. 
We confirmed the rotation of the envelope and disk in the \ce{^13CO} and \ce{C^17O} emission and found that their velocity gradients are the same as those of the outflow. 
Thus, we concluded that the velocity gradients in the low- and intermediate-velocity flows are due to the outflow rotation. 
Using observational outflow properties, we estimated the outflow launching radii to be $67.1-77.1\au$ for the low-velocity flow and $13.3-20.8\au$ for the intermediate-velocity flow. 
Although we could not detect the rotation in the jets due to the limited spatial resolution, we estimated the jet launching radii to be $(2.36-3.14) \times 10^{-2}\au$ using the observed velocity of each knots.
Thus, the jet is driven from the inner disk region. 
We could identify the launching radii of distinct velocity components within a single outflow with all the prototypical characteristics expected from recent theoretical works.

\end{abstract}

\keywords{Star formation(1569) ; Protostars(1302)}

\section{Introduction}
\label{sec:intro}

Protostellar outflows have been investigated for several decades as mass ejection phenomena in the early stages of star formation across various conditions in the Local Universe \citep[][and references there in]{Tokuda2022ApJL}.
Multiple velocity components, such as low-velocity wide-angle outflows and high-velocity collimated jets, are sometimes observed in protostellar outflows driven by a single protostar \citep{Snell1980ApJ}.
An easily comprehensible model illustrating these components was presented by \citet{Pudritznorman1986ApJ} from both observational and theoretical perspectives \citep[see Fig.~2 of][]{Pudritznorman1986ApJ}.
However, the driving mechanism behind these flows is still a subject of debate.
Two controversial scenarios, namely the $``$X-wind$"$ \citep{Shu1994ApJ} and $``$disk wind$"$ (\citealt{Blandfordpayne1982MNRAS}, more general treatment by \citealt{PeeletierPudritz1992ApJ}) have been proposed in theoretical studies to explain their origins.
In the X-wind scenario, only a high-velocity jet appears close to the protostar, which entrains the envelope (or infalling) gas and creates the low-velocity component \citep{Arce2007ppv}.
On the other hand, in the disk wind scenario, the flows originate from the entire disk region, resulting in several velocity components driven by different disk radii \citep[][and references therein]{Tsukamoto2022ppvii}.
To identify the driving mechanism of protostellar outflows, it is essential to investigate their launching region and angular momentum.
In the X-wind scenario, a small amount of angular momentum or very slow rotation velocity is expected for the low-velocity component, because the entrained gas has a small angular momentum \citep{Arce2007ppv}.
In contrast, the disk wind scenario predicts a large amount of angular momentum and fast rotation velocity for the low-velocity component, because it is driven by the outer disk region where the specific angular momentum is large \citep[e.g.,][]{Tomisaka2000ApJ,Machida2004,BanerjeePudritz2006ApJ}.

Three-dimensional numerical simulations of collapsing cloud cores have revealed that the low-velocity flow is launched from the outer disk region through the magneto-centrifugal acceleration mechanism, while the high-velocity flow emerges from the inner disk region due to the magnetic pressure gradient mechanism \citep[e.g.,][]{Machida2008ApJ,Tomida2010ApJ,Machida2019ApJ}.
These recent simulations have successfully reproduced the classical theoretical prediction of the disk wind, which explains the driving mechanisms of the flows.

Some observations can support such a theoretical picture of protostellar outflows \citep[][and references therein]{Pudritzray2019FrASS,Lee2020A&ARv}.
The multiple (velocity) components of outflows have been reported for a few objects \citep{Arce2013ApJ,Zapata2014MNRAS,Devalon2020aa,Lopes2023}.
\citet{Devalon2020aa} indicated the nested outflow, in which the outflow rotation was also detected.
However, since only a few objects showed multiple outflow components with rotation, more samples are needed to fully understand the driving of the protostellar outflow.

In addition to the studying the structures associated with the outflow morphology \citep[e.g.,][]{Arce2013ApJ}, it is crucial to observe the rotation of the outflow or jets and evaluate their angular momentum.
Quantitative analysis through observations will help to more accurately constrain or identify the driving mechanism.

Although outflows have been observed at various wavelengths, sub-millimeter/millimeter wavelength observations are particularly useful for understanding the detailed structure of protostellar outflows.
Recent ALMA observations have confirmed rotational signatures of both low-velocity outflow and high-velocty jet in several sources (e.g., CB26 \citeauthor{Laundhardt2009aa} \citeyear{Laundhardt2009aa}; TMC-1A \citeauthor{Bjerkeli2016Nature} \citeyear{Bjerkeli2016Nature}; Orion Source I \citeauthor{Hirota2017NatAs} \citeyear{Hirota2017NatAs}; HH30 \citeauthor{Louvet2018aa} \citeyear{Louvet2018aa}; DG Tauri B \citeauthor{Devalon2020aa} \citeyear{Devalon2020aa}; IRAS 16293-2422 A \citeauthor{Oya2021ApJ} \citeyear{Oya2021ApJ}; NGC1333 IRAS 4C \citeauthor{Zhang2018ApJ} \citeyear{Zhang2018ApJ}; HH212 \citeauthor{Tabone2017aa} \citeyear{Tabone2017aa}, \citeauthor{Lee2017NatAs} \citeyear{Lee2017NatAs}; OMC2/FIR 6b \citeauthor{Matsushita2021ApJ} \citeyear{Matsushita2021ApJ}; DG Tauri \citeauthor{Bacciotti2002ApJ} \citeyear{Bacciotti2002ApJ}; IRAS 18148-0440 \citeauthor{Oya2018ApJ} \citeyear{Oya2018ApJ}).
These observations support the disk wind mechanism proposed by \citet{Pudritznorman1986ApJ} as the driving mechanism of the low-velocity and high-velocity flow.
Especially, the rotation of both wide-angle and collimated flows have been reported in HH212. 
It should be noted that these rotational signatures were identified by different molecular line emissions, which mainly trace only strongly shocked gas, such as \ce{SO} and \ce{SiO}.
Robust examples of multiple, hierarchical outflow components and their rotations in \ce{CO} molecules that predominantly exist and more totally trace the bulk of the outflowing gas remain lacking.
To comprehensively understand the driving mechanism of these flows, it is crucial to detect the rotational motion and classify it into two or more components.

Our target, HH270mms1, is embedded within L1617 ($d = 405.7\,{\rm pc}$, \citealt{Tobin2020apj}), which is located in the northern part of the Orion B molecular cloud (typically $d = 407\,{\rm pc}$ \citealt{Kounkel2017ApJ,Kounkel2018AJ}).
The target was named in \citet{Tobin2020apj}; however, HH270mms1 is not directly associated with the Herbig-Haro object HH270 \citep{Garnavich1997ApJ}.
Other names for HH270mms1 are IRAS 05487+0255 \citep{Reipurth1991aa}, IRS 2 \citep{Davis1994ApJ}, and VLA 2 \citep{Rodriguez1998RMxAA}.
For this object, \citet{Tobin2020apj} resolved two continuum sources with the separation of $887\au$, and named each source HH270mms1-A and HH270mms1-B (hereafter mms1-A and mms1-B refers to as each continuum source and HH270mms1 refers to as a binary system).
The systemic velocity is $v_{\rm LSR} = 8.7\kms$ \citep{Choi2001ApJ,Sepuleveda2011aa}.
HH270mms1 was originally reported by \citet{Reipurth1991aa}, where infrared imaging revealed an extended nebulosity and a point-like source in the IRAS 05487+0255 field northwest of the HH110 launching point.
\citet{Davis1994ApJ} reported bright sources IRS 1, IRS 2 and reflection nebula in this field.
It should be noted that the \ce{CO} outflow from HH270mms1 was reported in \citet{Lee2000ApJ} and had contributed to construct the \textit{parabolic model} for the interpretation of the velocity structure exhibited in the protostellar outflow.
However, detailed observations of HH270mms1 have yet to be reported.

In this paper, we present evidence for nested outflows driven by HH270mms1, which serves as a suitable protostellar system to constrain the driving mechanism of the protostellar outflow. 
The properties of the observation toward HH270mms1 and data reductions are described in Section~\ref{sec:obs,reduc}.
The results of the outflow from HH270mms1, obtained using ALMA, are presented in Section~\ref{sec:results}.
We discuss the launching radii of nested outflows and the future identification of similar protostellar outflows in terms of small- and large-scale structural features in \S~\ref{sec:dis} and summarize our results in \S~\ref{sec:conclusion}.

\section{Observation and data reduction}
\label{sec:obs,reduc}

We analyzed ALMA archival data taken from two different projects, 2019.1.00086.S (PI: Ian Stephens) and 2015.1.00041.S (PI: John Tobin).
For the former project, the Band 7 observations were performed on  December 11 and December 20, 2019 with a 12-m array configuration with a base length of 15\,m -- 321\,m.
The total integration time toward HH270mms1 is 9.06\,minutes (543.6\,seconds).
The spectral windows (spws) include three continuum spectra and \ce{^12CO}($J$=\,3\,--\,2) spectrum.
The spws for continuum emissions were set as Frequency Division Mode (FDM), and the one of them includes \ce{C^17O}($J$=\,3\,--\,2) spectrum.
Although this project was set to obtain full polarization data, we focus on the dynamics around the protostar and do not use polarized emission (i.e.,\,Stokes Q,U,V) in this study.
The calibrator of Bandpass, Gain and Flux is J0552+0313, and Polarization calibrator is J0522--3627.
For the latter project, as reported in \citet{Tobin2020apj}, the Band 7 observations were carried out on September 4, 2016 and July 19, 2017. 
The total integration time is 0.9\,minutes (54\,seconds), and the configuration of 12-m Array are 15.1\,m -- 2.5\,km and 16.7\,m -- 3.7\,km in each observation date.
The Bandpass and Flux calibrator in the observation on  September 2016 is J0510+1800.
The Gain calibrator is J0552+0313.
The Bandpass calibrator in the observation on July 2017 is J0510+1800, the Flux calibrator is J0423--0120, and the Gain calibrator is J0552+0313.

\begin{table*}[hbt]
    \centering
    \caption{Parameters for imaging of each molecular line.}
    \label{tab:imaging}    
    \begin{tabular}{llccccc}
    \hline\hline
        molecule line & project code & spectral resolution\,($\kms$) & angular resolution & p.a.\,(deg) & rms\,(K) & robust parameter \\
        \hline
        \ce{^12CO}($J$=\,3\,--\,2) & 2019.1.00086.S & $0.25$ & $0\farcs57 \times 0\farcs73$ & $-68.1$ & 0.19 & $-1$ \\
        \ce{C^17O}($J$=\,3\,--\,2) & 2019.1.00086.S & $0.9$ & $0\farcs71 \times 0\farcs89$ & $-71.2$ & 0.038 & 0.5 \\
        \ce{^13CO}($J$=\,3\,--\,2) & 2015.1.00041.S & $0.5$ & $0\farcs34 \times 0\farcs36$ & $-57.7$ & 4.33 & 2.0 \\
        \hline
    \end{tabular}
\end{table*}

Calibration for raw data was performed using Common Astronomy Software Application \citep[CASA,][]{2022PASP..134k4501C}.
In the imaging process, we used \texttt{tclean} task with \texttt{h\"{o}gbom} deconvolver.
For both continuum and line imaging, the parameter of weighting is set \texttt{briggs}.
The robust parameter is set to $-1$ for \ce{^12CO} to obtain a higher angular resolution image with adequate sensitivity, although set to $0.5$ for \ce{C^17O}.
The \ce{^12CO} signal-to-noise ratio is still high enough to detect weaker emissions associated with high-velocity flow.
The robust parameter is set to $2.0$ for \ce{^13CO} to minimize the side-lobe artifact due to a short integration time.
In addition, for \ce{^13CO} imaging, we extract baseline $>\,30\,k\lambda$ and select the parameter for \texttt{uvtaper} as $300\,k\lambda$.
The synthesized beam and noise level of each molecular line are listed in Table~\ref{tab:imaging}.

\section{Results}
\label{sec:results}

\subsection{Layered flows from HH270mms1-A} 
\label{res:12CO}

\begin{figure*}[htb]
    \centering
    \includegraphics[width=\textwidth]{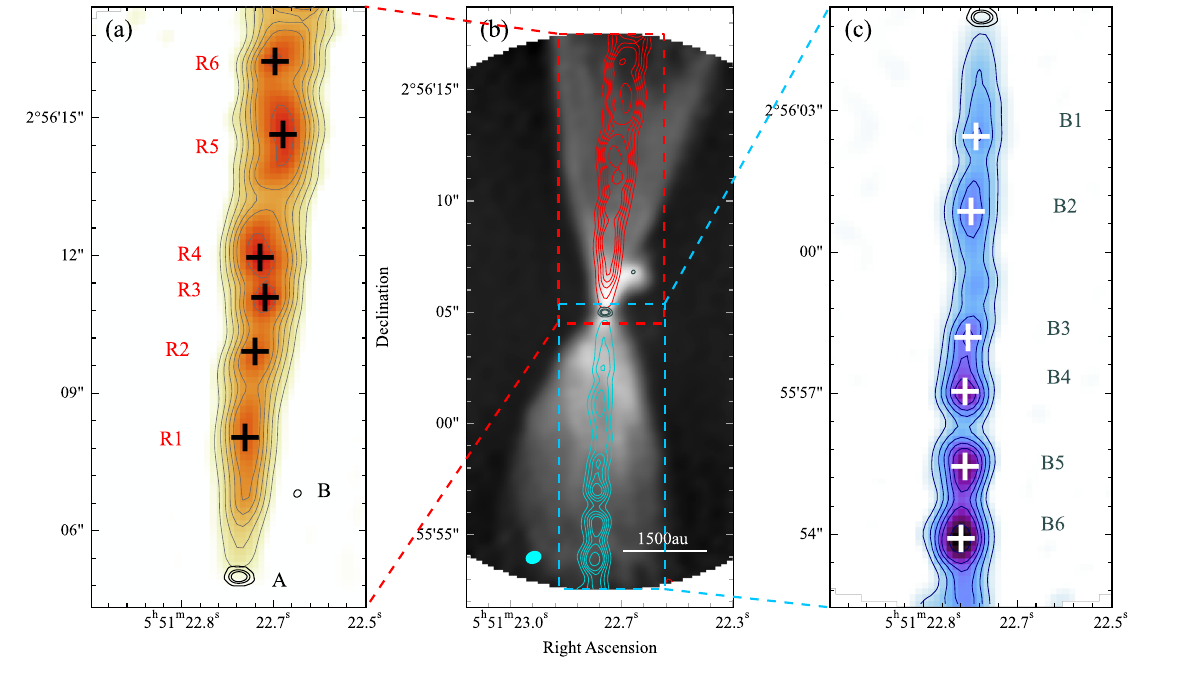}
    \caption{
    (a) \ce{^12CO} integrated intensity map of the redshifted jet (color and gray contours) in the velocity range described in the bottom row of  Table~\ref{tab:vboundary_range}. 
    (b) Redshifted (red contours) and blueshifted (cyan contours) jets superimposed on the low-velocity outflow (gray color scale, integrated in the velocity range as in \citealt{Lee2000ApJ}).
    The cyan ellipse in the bottom left corner represents the synthesized beam size.
    (c)  \ce{^12CO} integrated intensity of the blueshifted jet (color and blue contours) in the velocity range described in the bottom row of Table~\ref{tab:vboundary_range}.
    The identified knots (the peak position in the integrated intensity map) are indicated by black and white crosses in panels (a) and (c) and labeled as R1--R6 and B1--B6.
    In the panels (a)-(c), the black contours indicate the continuum emission with the levels of $[10,50,100]\,\sigma, 1\sigma = 0.36\,\textrm{mJy\,beam}^{-1}$ and show the positions of HH270mms1-A and HH270mms1-B.
    The contour levels of \ce{^12CO} integrated intensity are $[10,20,30,40,50,75,100]\,\sigma, 1\sigma = 1.19\Kdeg\kms$ for redshifted component, $[10,20,30,40,50,75,100]\,\sigma, 1\sigma = 1.37\Kdeg\kms$ for blueshifted component.
    }
    \label{fig:knots}
\end{figure*}

\begin{table}[htb]
    \centering
    \caption{Velocity range of the main components of the layered outflow from HH270mms1-A.}
    \label{tab:vboundary_range}
    \begin{tabular}{lrr}
        \hline\hline
         & red\,($v_{\rm LSR},\kms$)& blue\,($v_{\rm LSR},\kms$)\\
        \hline
        low & 8.7 -- 16.0 & 3.0 -- 8.7\\
        intermediate & 16.5 -- 34.5 & $-16.0$ -- 3.0\\
        high & 46.0 -- 70.0 & $-65.0$ -- $-40.0$\\
        \hline
    \end{tabular}
\end{table}

\begin{table*}[htb]
\centering
    \centering
    \caption{Physical properties of identified knots for the refshifted jet.}
    \label{tab:redknots}
    \begin{tabular}{lcccc}
        \hline\hline
         & projected distance\,(au) & $v_{\rm LSR}\,(\kms)$  & $t_{\rm dyn}\,(\yr)$ & $\Delta t_{\rm dyn}\,(\yr)$\\
        \hline
        R1 & $1.2 \times 10^3$ & $63.1$ & $84.4$ & $84.4$ \\
        R2 & $1.9 \times 10^3$ & $65.6$ & $128$ & $44.1$ \\
        R3 & $2.4 \times 10^3$ & $62.8$ & $172$ & $44.1$ \\
        R4 & $2.8 \times 10^3$ & $62.0$ & $197$ & $24.9$ \\
        R5 & $3.8 \times 10^3$ & $63.9$ & $260$ & $62.5$ \\
        R6 & $4.6 \times 10^3$ & $63.3$ & $332$ & $63.8$ \\
        \hline
    \end{tabular}
\end{table*}

\begin{table*}[htb]
    \centering
    \caption{Physical properties of identified knots for the blueshifted jet.}
    \label{tab:blueknots}
    \begin{tabular}{lcccc}
        \hline\hline
         & projected distance\,(au) & $v_{\rm LSR}\,(\kms)$ & $t_{\rm dyn}\,(\yr)$ & $\Delta t_{\rm dyn}\,(\yr)$\\
        \hline
        B1 & $1.0 \times 10^3$ & $-52.2$ & $63.4$ & $63.4$ \\
        B2 & $1.7 \times 10^3$ & $-51.5$ & $107$ & $43.9$ \\
        B3 & $2.3 \times 10^3$ & $-44.3$ & $167$ & $59.3$ \\
        B4 & $2.8 \times 10^3$ & $-44.5$ & $198$ & $31.0$ \\
        B5 & $3.2 \times 10^3$ & $-43.7$ & $237$ & $38.8$ \\
        B6 & $3.8 \times 10^3$ & $-44.1$ & $277$ & $40.2$ \\
        \hline
    \end{tabular}
\end{table*}

\begin{figure*}[htb]
    \centering
    \includegraphics[width=\textwidth]{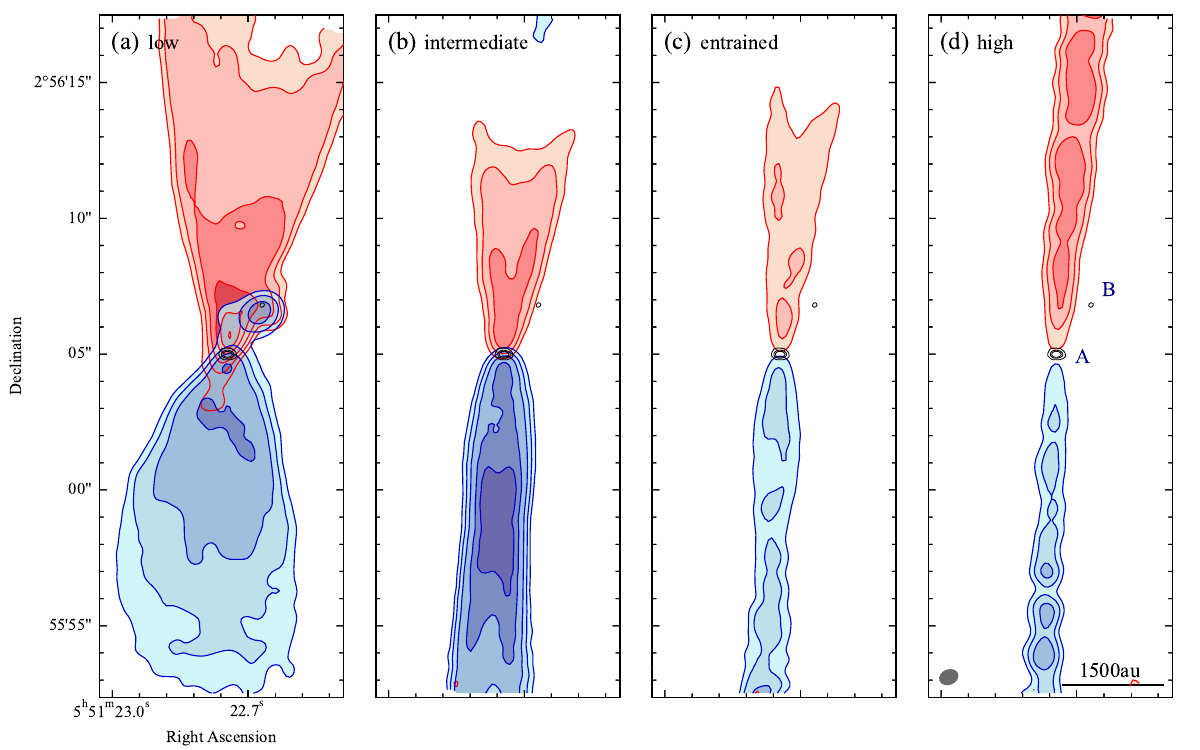}
    \caption{
    \ce{^12CO} integrated intensity maps for (a) low-velocity, (b) intermediate-velocity, (c) entrained, and (d) high-velocity flows.
    The contour levels for redshifted and blueshifted components are $[10,25,50,100,150,200] \sigma, 1\sigma = 1.34\Kdeg\kms$ and $1.92\Kdeg\kms$ in panel (a), $1.47\Kdeg\kms$ and $1.55\Kdeg\kms$ in panel (b), $1.44\Kdeg\kms$ and $1.85\Kdeg\kms$ in panel (c), $1.82\Kdeg\kms$ and $1.36\Kdeg\kms$ in panel (d), respectively.
    }
    \label{fig:4fig}
\end{figure*}

\begin{figure*}[htb]
    \centering
    \includegraphics[width=\textwidth]{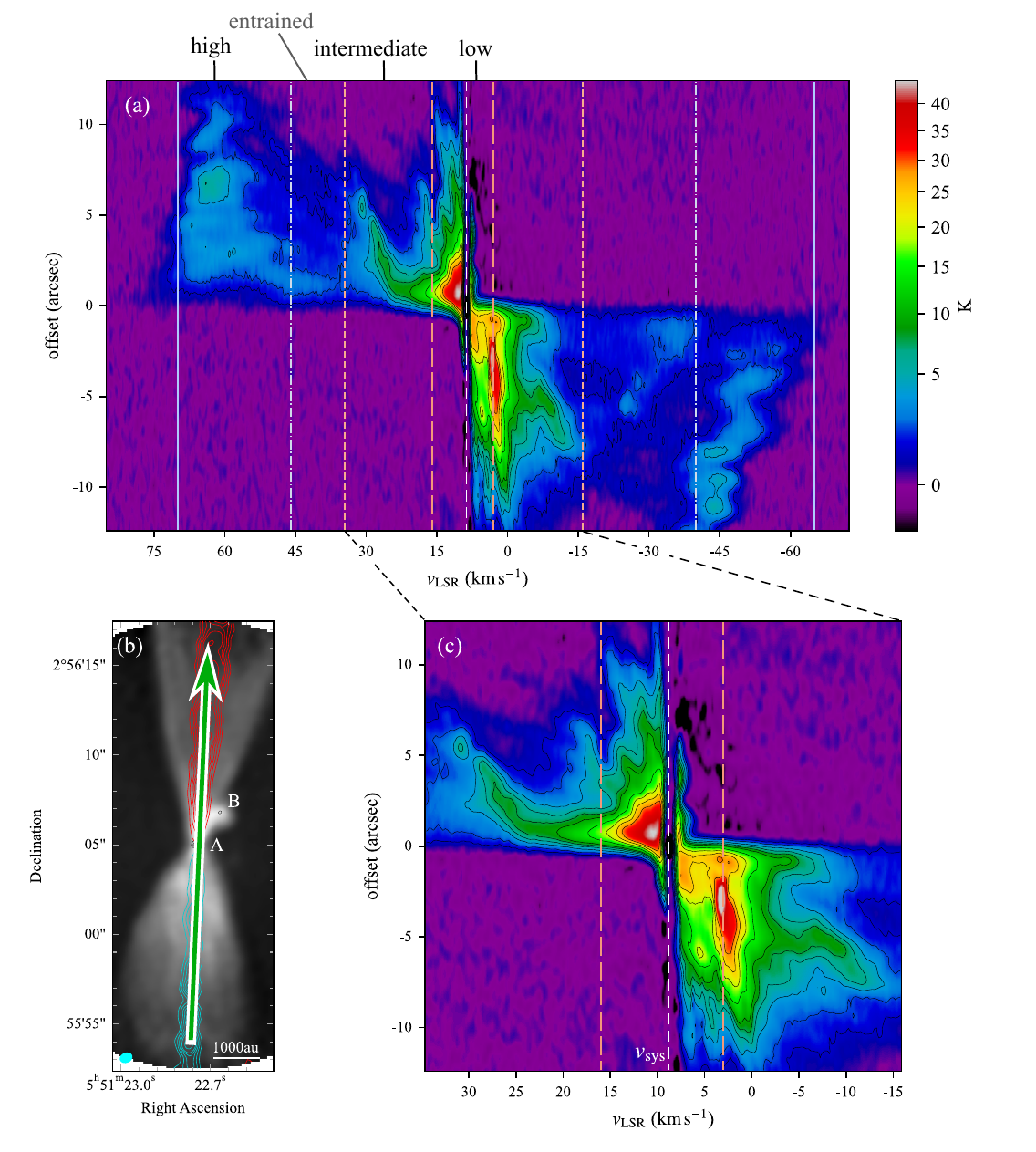}
    \caption{
    (a) Position-Velocity (PV) diagram along the outflow axis. The black contours indicate the \ce{^12CO} brightness temperature with the contour levels of $[5, 10, 20, 30, 40, 50, 75, 100, 125, 150]\,\sigma,1\sigma = 0.19\Kdeg$. 
    The $x$-axis is described as the velocity of local standard of rest (LSR).
    The vertical lines indicate the boundaries of each flow component as descried in Table~\ref{tab:vboundary_range} and the dashed white line shows the systemic velocity of HH270mms1, $v_\textrm{sys} = 8.7\kms$.
    (b) Integrated intensity map as in Figure~\ref{fig:knots}(b). The green arrow indicates the direction for the slice of PV diagram in panel (a). The positions of the continuum emissions are labeled as A and B for mms1-A and mms1-B, respectively.
    (c) Zoomed image of the lower velocity range (low- and intermediate-velocity outflows) in panel (a). The color scale and contour levels are the same as in panel (a).
    }
    \label{fig:pv_parallel}
\end{figure*}

Figure~\ref{fig:knots} clearly illustrates the connection between the bipolar jets and mms1-A, indicating that the jets are driven from mms1-A.
We could also confirm that both the redshifted and blueshifted jets contains six bright knots (Figure~\ref{fig:knots}).
We refer to these knots as R1, R2, \dots, R6 in the redshifted jet, while they are denoted as B1, B2, \dots, B6 in the blueshifted jet, as labeled in the panels (a) and (c) of Figure~\ref{fig:knots}.
Similar structures, characterized by several knots within the jets, have been observed in other protostellar systems.
It is considered that the presence of these knots has been attributed to episodic mass ejection phenomena as described in previous studies \citep{Wang2019ApJ}, and which is discussed in \S\ref{subsec:jet}.
It should be noted that, for this object, \citet{Lee2000ApJ} reported only the low-velocity component mentioned above.
In contrast to \citet{Lee2000ApJ}, we have, for the first time, detected well-collimated high-velocity component (hereafter, jet) driven from mms1-A and surrounded by wide-angle low-velocity flow.
Considering the morphological features of the integrated intensity maps shown in Figure~\ref{fig:4fig}(a) and (d), and the arcmin-scale low-velocity outflow reported in \citet{Lee2000ApJ}, both the jet and the low-velocity outflow should extend over the Field of View (FOV) with the present ALMA observation. 

Figure~\ref{fig:pv_parallel} shows the Position-Velocity (PV) diagram cutting along the outflow axis, in which the cutting direction is shown in Figure~\ref{fig:pv_parallel}(b).
The projected maximum velocity of the jets is $\sim 70.0\kms$ for the redshifted component and $\sim -65.0\kms$ for the blueshifted component.
Adopting the disk inclination angle $i$ of mms1-A with respect to the plane of the sky $i=51.1\deg$ inferred from our high-resolution continuum image, we estimated the deprojected velocity of the jets assuming that the jets propagate in the normal direction of the circumstellar disk, with the maximum velocity of $\sim 80 \kms$ from the systemic velocity of mms1-A.
It is also possible to estimate the dynamical timescale of mass ejections $t_{\rm dyn}$, in which the projected distance between each knot and the protostar $l_{\rm out}$ is divided by the velocity of each knot $v_{\rm LSR}$ (listed in Table~\ref{tab:redknots} and \ref{tab:blueknots}) as
\begin{equation}
    t_{\rm dyn} = \dfrac{l_{\rm out}}{(v_{\rm LSR}\tan{i})},  
\label{eq:t_dyn}
\end{equation}
where $\tan{i}$ is the correction factor for the inclination angle of the jets.
We estimated the time interval between the knots to be approximately $\Delta t_{\rm dyn} \sim 47.9$ years for the redshifted jet and $\Delta t_{\rm dyn} \sim 42.6$ years for the blueshifted one.
Table~\ref{tab:redknots} and \ref{tab:blueknots} provide the physical properties of each knot in the redshifted and blueshifted jet, respectively.
The deprojected maximum velocity of the jets is comparable to the velocities of previously reported EHV flows \citep[e.g.,][]{Hirano2010ApJ, Hull2016ApJ}.

In the PV diagram (Figure~\ref{fig:pv_parallel}(a) and (c)), two pairs of velocity gradients in lower velocity range and plateaued velocity components in higher velocity range are clearly exhibited.
Based on the PV diagram and channel maps, we categorized those driven from mms1-A into three velocity components: low-velocity, intermediate-velocity, and high-velocity (jets) as shown in Figure~\ref{fig:4fig}.
The velocity ranges for each component are listed in Table~\ref{tab:vboundary_range}.
Note that the velocity range of the low-velocity component is set to be the same as reported in \citet{Lee2000ApJ}.
In addition to the three components with strong \ce{^12CO} emission, we also observed weak emission in the velocity range of $v_\textrm{LSR} = -40.0\kms$ to $-16.0\kms$ and $34.5\kms$ to $46.0\kms$.
The emission in this velocity range is considerably weaker compared to the other three components, and we cannot observe a clear accelerating pattern (Figure~\ref{fig:pv_parallel}(a)).
Hence, we expect that the flow in this velocity range is entrained by the bipolar jets, and we refer to it as the entrained component.
The vertical lines in Figure~\ref{fig:pv_parallel}(a) and (c) indicate the velocity boundaries of each component (see also Table~\ref{tab:vboundary_range}).
In Figure~\ref{fig:pv_parallel}(c), the velocities in the low- and intermediate-velocity components increase with distance from mms1-A, resembling the \textit{ Hubble law} in the context of protostellar outflows. 
On the other hand, we did not observe the \textit{Hubble law}-like acceleration in the jets.
Instead, we found the oscillation along the outflow axis in the range of $-45 \kms \lesssim v_{\rm LSR} \lesssim -60\kms $ for the blueshifted jet as shown in Figure~\ref{fig:pv_parallel}(a), suggesting episodic mass ejection \citep[e.g.,][]{Stonenorman1993ApJ}.
Each local emission peak in the jets roughly corresponds to each knot seen in Figure~\ref{fig:knots}.

Figure~\ref{fig:4fig} shows \ce{^12CO} integrated intensity maps of each velocity components defined above.
The overall opening angle in the low-velocity components are wider than the higher velocity components.
The configuration of the higher velocity component surrounded by the lower velocity component is consistent with the numerical simulations \citep[e.g.,][]{Machida2008ApJ}.
In Figure~\ref{fig:4fig}(a), the excess of \ce{^12CO} emission of the low-velocity component exists around mms1-B.
However, considering the absence of the higher velocity components, mms1-B does not seem to affect the large scale structures of the bipolar outflow.
The entrained components we have defined and shown in Figure~\ref{fig:4fig}(c) have less distinct structures than the components seen in panels (b) and (d), and their opening angles are intermediate in magnitude between them.
Moreover, the contour levels in Figure~\ref{fig:4fig}(c), which represents the  \ce{^12CO} integrated intensity, are also lower than those in panels (b) and (d).

\subsection{Outflow physical quantities}
\label{sec:outflowquantities}

\begin{table*}[htp]
    \centering
    \caption{Physical quantities of each velocity component.}
    \begin{tabular}{lrrrr}
    \hline\hline
    component & $v_{\rm typ}$ ($\kms$)& $\dot{M}_{\rm out}$ ($\Msun \yr^{-1}$) & $F_{\rm out}$ ($\Msun \kms \yr^{-1}$) & $L_{\rm out}$ ($\erg\yr^{-1}$)\\
    \hline
    low & $5.6$ & $5.0 \times 10^{-8}$ & $2.8 \times 10^{-7}$ & $3.1 \times 10^{37}$ \\
    intermediate & $22.4$ & $1.4 \times 10^{-7}$ & $3.1 \times 10^{-6}$ & $1.4 \times 10^{39}$ \\
    high & $87.7$ & $1.0 \times 10^{-7}$ & $5.5 \times 10^{-6}$ & $6.0 \times 10^{39}$ \\
    \hline
    total & & $2.9 \times 10^{-7}$ & $8.9 \times 10^{-6}$ & $7.4 \times 10^{39}$ \\
    \hline
    \end{tabular}
    \label{tab:outflowquantities}
\end{table*}

In this subsection, we derive the outflow physical quantities, outflow mass loss rate $\dot{M}_{\rm out}$, outflow force $F_{\rm out}$ and kinetic luminosity $L_{\rm out}$, integrating the four velocity components shown in Figure~\ref{fig:4fig}.
Assuming the local thermodynamical equilibrium~(LTE) and optically thin \ce{CO} emission, we estimated the outflow mass as 
\begin{equation}
    M_{\rm out} = \mu m_{\rm H} X_{\rm CO}^{-1} \Omega d^2 N_{\rm CO},
\label{eq:masscolumn}
\end{equation}
where $\mu=2.8$ is the mean molecular weight per hydrogen molecules, $m_{\rm H}$ is the hydrogen atom mass, $\Omega$ is the total solid angle, and $d$ is the distance to the source, 405.7\,pc. We assumed the CO abundance ratio with respect to H$_2$, $X_{\rm CO}=10^{-4}$ \citep[e.g.,][]{Frerking1982ApJ}, and excitation temperature, $T_{\rm ex}=40\Kdeg$.
Note that for molecular outflow, the excitation temperature of \ce{^12CO}($J$=\,3\,--\,2) could be in the range of $T_{\rm ex} \sim 20 - 50 \Kdeg$ \citep[e.g.,][]{Takahashi2008ApJ,Ginsburg2011MNRAS,Dunham2014ApJ}.
However, as shown in Figure A1 of \citet{Ginsburg2011MNRAS}, the derived column density does not significantly depend on the excitation temperature in the range ($T_{\rm ex} = 20 - 50 \Kdeg$).
The column density has a local minimum around $T_{\rm ex} \simeq 40 \Kdeg$.
The difference in column density is within 20\% with the change of $T_{\rm ex}$ in the range of $20$ to $50\Kdeg$. 
Thus, although we have adopted an excitation temperature of $T_{\rm ex} = 40 \Kdeg$, changing the excitation temperature does not significantly change our results. 
In addition, in order to define the outflow emission boundaries, a $10 \sigma$ threshold is set for each integrated intensity map in Figure~\ref{fig:4fig}.

For each velocity component, we determined the dynamical timescale with equation~(\ref{eq:t_dyn}) using the outflow length $l_{\rm out}$ and typical velocity $v_{\rm typ}$.
We used the averaged dynamical timescale for high-velocity components (Tables~\ref{tab:redknots} and \ref{tab:blueknots}).
Using the mass, typical velocity and dynamical timescale of each outflow component, we derived the outflow mass loss rate,
\begin{equation}
\dot{M}_{\rm out} = \dfrac{M_{\ce{out}}}{t_{\rm dyn}},
\end{equation}
the outflow momentum flux
\begin{equation}
F_{\rm out}= \dfrac{M_{\ce{out}}v_{\rm typ}}{t_{\rm dyn}},
\end{equation}
and the outflow kinetic luminosity 
\begin{equation}
L_{\rm out}= \dfrac{1}{2} \dfrac{M_{\ce{out}}v_{\rm typ}^2}{t_{\rm dyn}}.
\end{equation}
The derived values for each component are listed in  Table~\ref{tab:outflowquantities}.
We considered all velocity components (low-, intermediate- and high-velocity components) of the outflow to derive the total outflow physical quantities, $\dot{M}_{\rm out}=2.9 \times 10^{-7} \Msun \yr^{-1}$, $F_{\rm out}=8.9 \times 10^{-6} \Msun\, \kms \yr^{-1}$, and $L_{\rm out}=7.4 \times 10^{39} \erg\, \yr^{-1}$. 
These values are comparable to those shown in previous observational studies of Class~0/I objects \citep[e.g.,][]{Arce2006ApJ,Feddersen2020ApJ,Hsieh2023ApJ}. 

Finally, we comment on the optically thin assumption of the outflow. 
The optically thin assumption of \ce{CO} is not always satisfied, especially in the low-velocity outflow component \citep{Hsieh2023ApJ}.
Thus, the quantities listed in Table~\ref{tab:outflowquantities} may give a lower limit.

\subsection{Velocity gradient perpendicular to outflow axis and rotation signature}
\label{res:rot}

\begin{figure*}[htb]
    \centering
    \includegraphics[width=\textwidth]{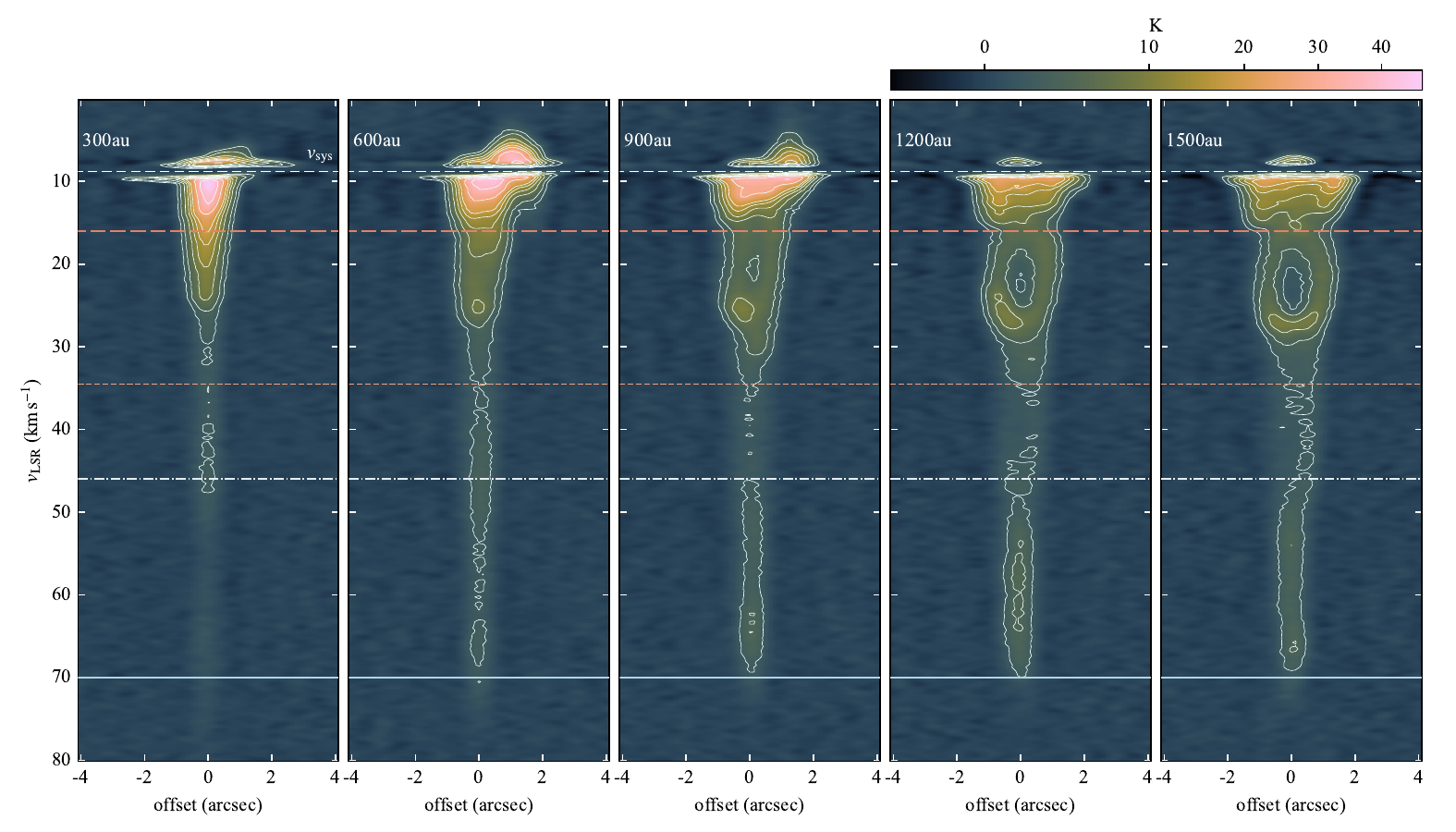}
    \caption{
    PV diagrams of the \ce{^12CO} redshifted outflow perpendicular to the outflow axis, at the projected distance every 300\,au from HH270mms1-A.
    The left-to-right direction of each PV diagram corresponds to the east-to-west direction of the field of view.
    The contour levels are $[10,20,40,60,80,100,150,200]\,\sigma, 1\sigma = 0.19\Kdeg$.
    The horizontal lines represent the velocity boundaries of each outflow component and the systemic velocity as in Figure~\ref{fig:pv_parallel}.
    }
    \label{fig:pvmaps}
\end{figure*}

\begin{figure*}[htb]
    \centering
    \includegraphics[width=\textwidth]{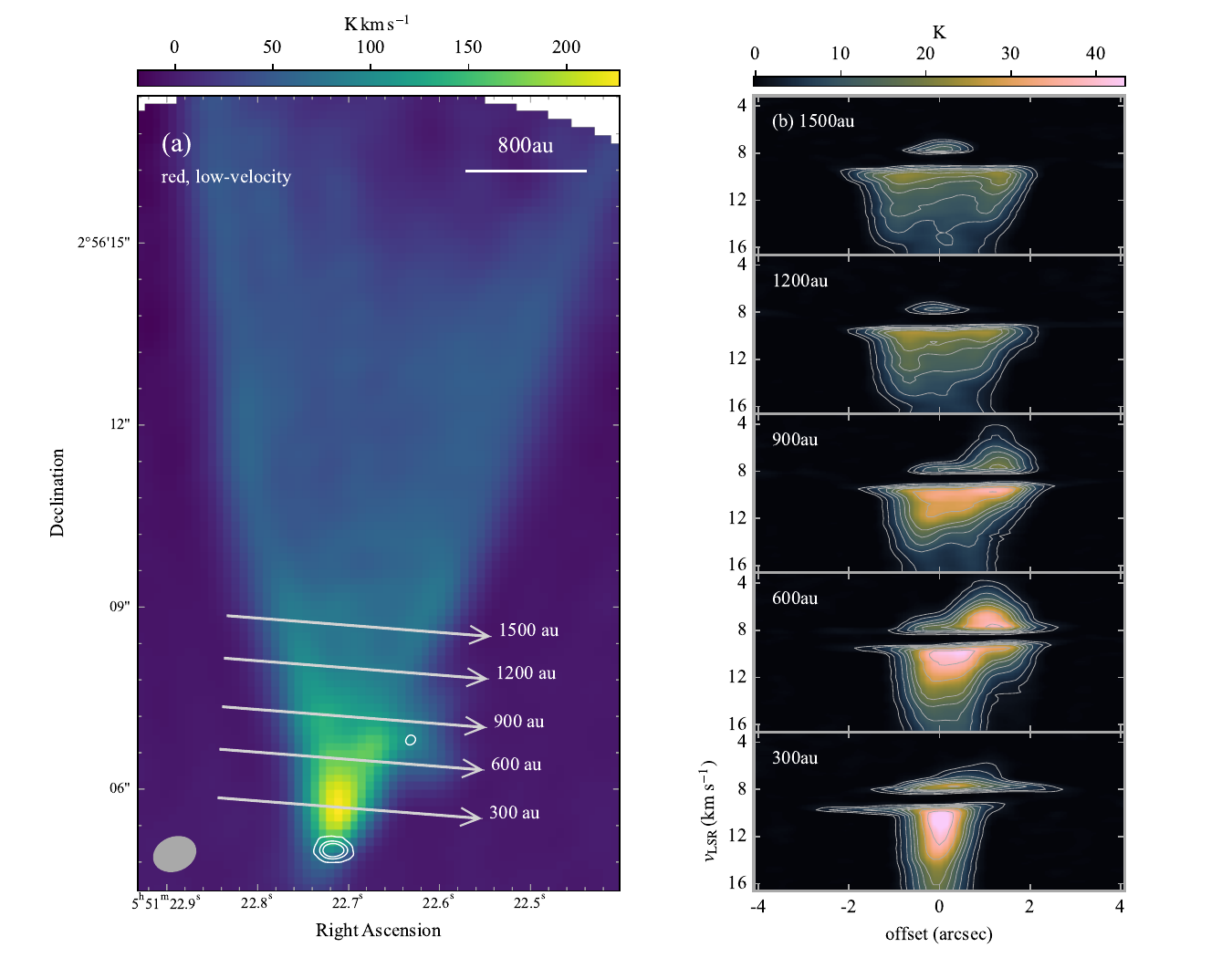}
    \caption{
    (a) Integrated intensity of the northern part of the outflow as shown in Figure~\ref{fig:4fig}(a) (color) and the guide lines for PV diagrams (gray solid arrows) every $300\,\au$ intervals from HH270mms1-A.
    The gray ellipse indicates the synthesized beam size (Table~\ref{tab:imaging}).
    The white contours indicate 0.87\,mm continuum emission as in Figure~\ref{fig:knots}. 
    (b) Zoomed PV diagrams of Figure~\ref{fig:pvmaps} focusing on the low-velocity components. 
    The contour levels is the same as in Figure~\ref{fig:pvmaps}.
    }
    \label{fig:pv_low}
\end{figure*}

\begin{figure*}[htb]
    \centering
    \includegraphics[width=\textwidth]{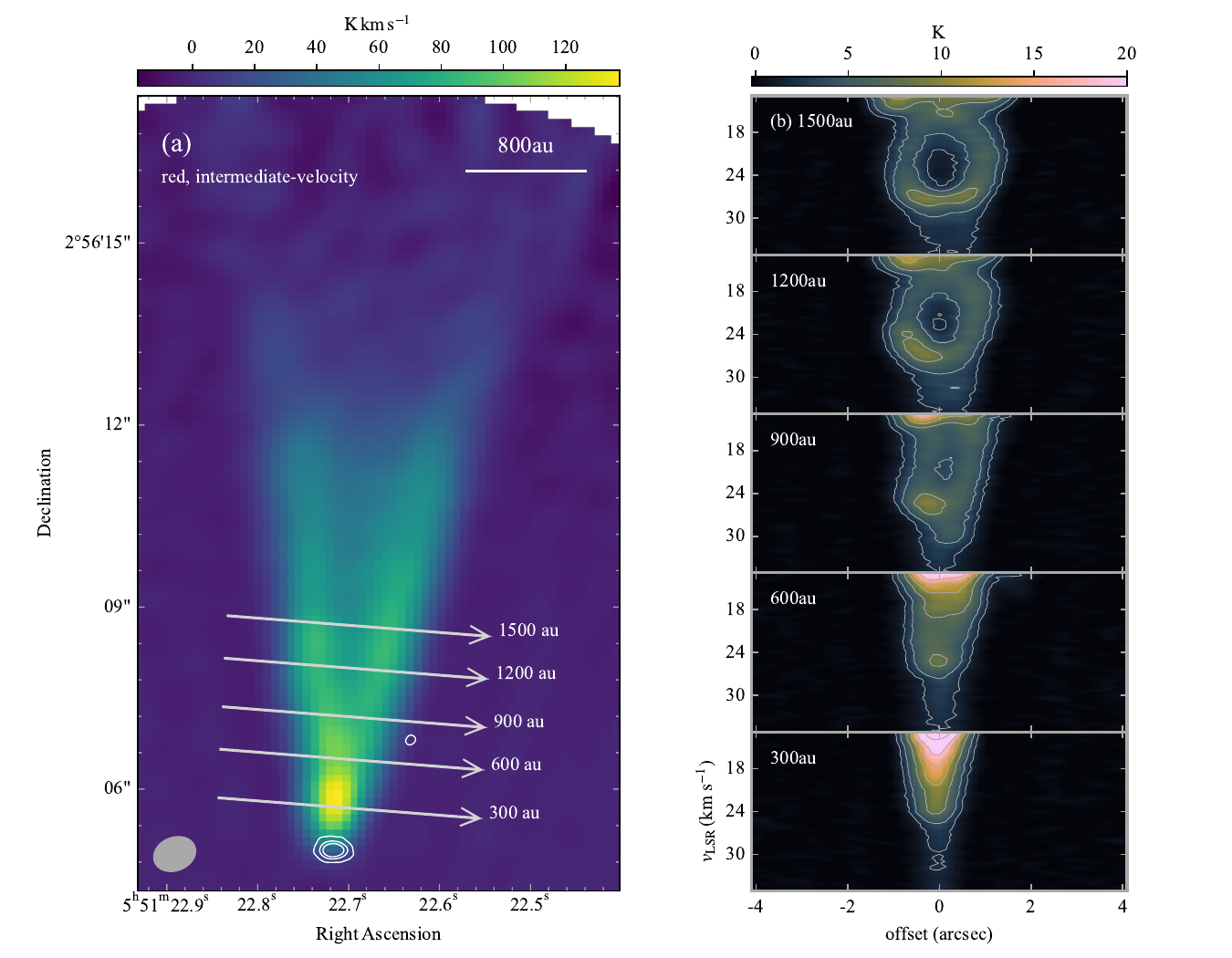}
    \caption{
    Same as Figure~\ref{fig:pv_low} but focusing on intermediate velocity components.
    }
    \label{fig:pv_intermediate}
\end{figure*}

\begin{table*}[htb]
    \centering
    \caption{Observed quantities of the low- and intermediate-velocity components derived with Figures~\ref{fig:pv_low} and \ref{fig:pv_intermediate}, and estimated outflow physical quantities.}
    \label{tab:rot}
    \begin{tabular}{lccrclccr}
        \hline\hline
        \multicolumn{4}{c}{Observed quantities} && \multicolumn{4}{c}{Derived quantities}\\
        \cline{1-4} \cline{6-9}
        $z$\footnote{projected distance from mms1-A.} & $v_{p,\infty}$\footnote{poloidal velocity of outflow subtracted the systemic velocity of mms1-A, $v_{\rm sys} = 8.7\kms$.} & $v_{\phi,\infty}$ & $r_{\infty}$ && $j = v_{\phi,\infty} r_{\infty}$ & $r_0$ & $r_A$ & $\lambda\equiv r_a/r_0$ \\
        (au) & ($\rm{km\,s}^{-1}$) & ($\rm{km\,s}^{-1}$) & (au) && ($\rm{cm}^2\,\rm{s}^{-1}$)& (au) & (au) & -\\
        \cline{1-4} \cline{6-9}
        \multicolumn{9}{c}{low-velocity}\\
        \hline
        900 & 6.17 & 1.96 & 365.1 && $1.07\times 10^{21}$ & 67.1 & $1.63\times 10^2$ & 2.42\\
        1200 & 5.17 & 1.22 & 507.1 && $9.25\times 10^{20}$ & 77.1 & $1.68\times 10^2$ & 2.18\\
        1500 & 5.49 & 1.09 & 567.9 && $9.27\times 10^{20}$ & 71.3 & $1.58\times 10^2$ & 2.22\\
        \hline
        \multicolumn{9}{c}{intermediate-velocity}\\
        \hline
        900 & 22.6 & 2.83 & 300.2 && $1.27\times 10^{21}$ & 13.3 & $5.26\times 10^1$ & 3.95\\
        1200 & 20.6 & 3.40 & 405.7 && $2.07\times 10^{21}$ & 20.8 & $9.39\times 10^1$ & 4.51\\
        1500 & 23.9 & 3.01 & 417.8 && $1.89\times 10^{21}$ & 16.1 & $7.38\times 10^1$ & 4.60\\
        \hline
    \end{tabular}
\end{table*}

Figure~\ref{fig:pvmaps} shows the \ce{^12CO} PV diagram perpendicular to the redshifted outflow axis (or outflow propagation direction).
We made the PV diagram every 300 au, which is the projected distance from mms1-A.
In order to focus on each velocity component, Figures~\ref{fig:pv_low}(b) and \ref{fig:pv_intermediate}(b) show the zoomed images of Figure~\ref{fig:pvmaps}.
The gray arrows in Figures~\ref{fig:pv_low}(a) and \ref{fig:pv_intermediate}(a) 
indicate the direction of the slice of the PV diagram (Figures~\ref{fig:pvmaps}, \ref{fig:pv_low}(b) and \ref{fig:pv_intermediate}(b)).
Figure~\ref{fig:pv_low}(a) shows that the width of low-velocity component broadens with distance from mms1-A.
In Figure~\ref{fig:pv_low}(b), the excess of blueshifted emission at $600\au$ and $900\au$ traces the emission associated with mms1-B as seen in Figure~\ref{fig:4fig}(a).
In Figure~\ref{fig:pv_intermediate}, a ring-like structure can be seen in the PV diagram of the intermediate-velocity component, particularly far from mms1-A. 
Furthermore, the ring at 1200 au seems to be distorted, which can be attributed to the velocity gradient perpendicular to the outflow axis or along the line of sight.
Similar ring-like structures have been reported in some edge-on protostellar outflows \citep[e.g.,][]{Hirota2017NatAs,Oya2018ApJ}. 
This distorted ring-like structure in the PV diagram can be explained by a rotating and expanding shell and is discussed in \S\ref{subsec:identify}.
The PV diagrams of the low-velocity component exhibit the same velocity gradients as in the intermidiate-velocity components, and similar structure was reported in \citet{Zapata2015ApJ}.
Considering that the asymmetric velocity structures observed in Figures~\ref{fig:pv_low} and \ref{fig:pv_intermediate} are due to the rotation of outflow, we calculated the physical properties of both the low- and intermidiate-velocity components, which are listed in Table~\ref{tab:rot}.
In this study, we only derived the redshifted, northern outflow parameters because the blueshifted, southern one does not clearly show asymmetric velocity distribution along the outflow axis.
It is possible that the asymmetric component of the blue-shifted flow is obscured by some observational effect, but there is currently no clear explanation.
However, since the two velocity components generally show similar structures as a whole (Figures~\ref{fig:pvmaps} and \ref{apfig:pv_blue}), we would like to point out that the discussion based on the derivation of the red-shifted component may also be applicable to the blue-shifted component.
We briefly discuss the rotation of the blue-shifted component in \S\ref{subsec:r_l}.

In contrast to the lower velocity components, velocity gradients could not be confirmed in the high-velocity components (Figure~\ref{fig:pvmaps}).
The reason for this can be due to insufficient spatial resolution.
As seen in Figures~\ref{fig:4fig} and \ref{fig:pvmaps}, the synthesized beam size in the \ce{^12CO} observation is comparable to the width of high-velocity component. 

\subsection{CO isotope molecular lines and rotation of mms1-A}
\label{res:13COetC17O}

\begin{figure*}[htb]
    \centering
    \includegraphics[width=\textwidth]{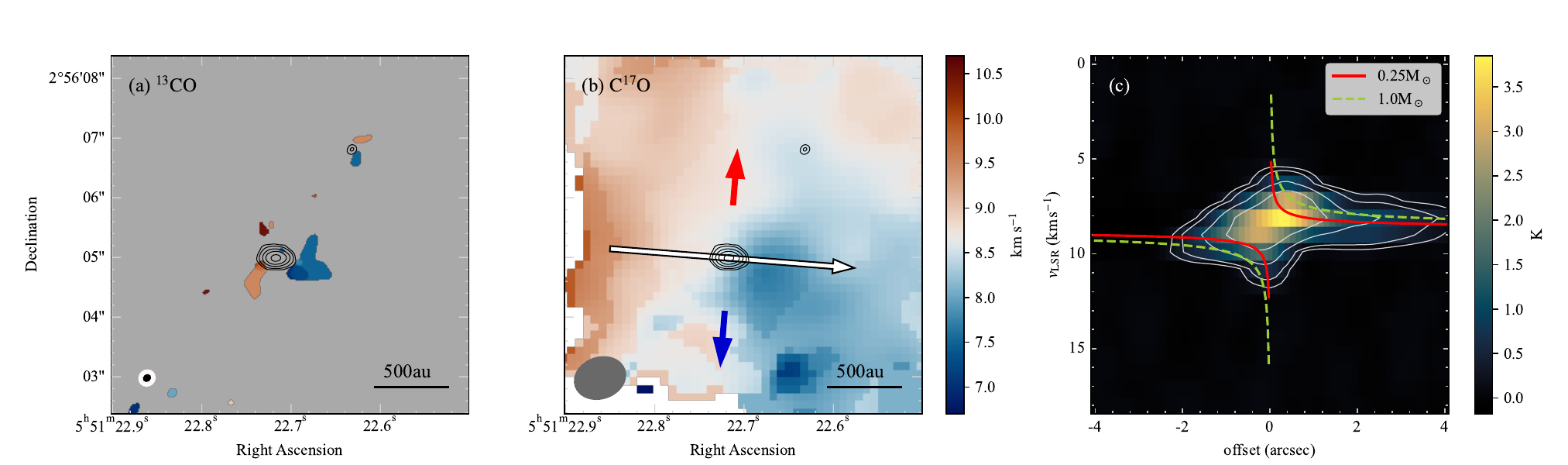}
    \caption{
    (a) Mean velocity map (moment 1) of the \ce{^13CO} emission. 
    The black contours indicate the continuum emission with the contour levels of $[10,20,50,100,200]\,\sigma, 1\sigma = 0.36\,\textrm{mJy\,beam}^{-1}$.
    The white ellipse and the black ellipse in the bottom left corner indicate the synthesized beam size of the continuum and \ce{^13CO} image.
    (b) Mean velocity map of the \ce{C^17O} emission. The white arrow indicates the direction for the PV diagram shown in panel (c).
    The black contours are the same as those in panel (a).
    The gray ellipse indicates the synthesized beam size of \ce{C^17O} image.
    The red and blue arrows show the direction of outflow and jet.
    Mean velocity maps were made with the intensity higher than $3\sigma$ (panel a) and $5\sigma$ (panel b) level, respectively.
    (c)  \ce{C^17O} PV diagram along the major axis of mms1-A continuum emission.
    Red solid line and green dashed line correspond to the Keplerian rotation profile of the mass of $0.25\,\Msun$ and $1.0\,\Msun$, respectively.
    The white contour levels are $[5, 10, 25, 50]\,\sigma, 1\sigma = 0.038\,{\rm K}$.}
    \label{fig:13COC17O_pv}
\end{figure*}

To confirm the velocity gradient of the outflow due to rotation, it is necessary to observe the rotation of the disk located at the foot of the outflow.
The rotation axis of mms1-A disk was unclear because the previous observations did not have a sufficient spatial resolution to trace rotational motion around the protostar or focused on only the continuum emission.
The rotation direction of the outflow should be the same as that of the circumstellar disk as long as the outflow is directly driven by the circumstellar disk \citep{Blandfordpayne1982MNRAS}. 
Since the density of the outflow launching region should be high, \ce{^13CO} and \ce{C^17O} have the potential to trace the kinematics of the gas in such a region.
Figure~\ref{fig:13COC17O_pv}(a) and (b) show mean velocity maps obtained from \ce{^13CO} and \ce{C^17O} line emission, respectively. 

Due to the limited sensitivity, \ce{^13CO} emissions can be only detected in the vicinity of the continuum emission. 
However, the rotation direction is consistent with the expected rotation direction of the outflow, supporting the presence of rotational motion in the outflow.
Although the sensitivity of the \ce{^13CO} is limited, the emission from the \ce{C^17O} is sufficient to detect the entire region surrounding mms1-A. 
Figure~\ref{fig:13COC17O_pv}(b) clearly shows the velocity gradient around mms1-A continuum emission. 
For the envelope scale, the gas on the eastern side moves towards the far side, while on the western side, it moves towards the near side.
In addition, the direction of the rotation (or rotation axis), which traces the low-velocity part (white region in Figure~\ref{fig:13COC17O_pv}(b), i.e., near the systemic velocity), is roughly aligned with the propagation direction of the outflow. 

Figure~\ref{fig:13COC17O_pv}(c) shows the \ce{C^17O} PV diagram along the major axis of mms1-A continuum emission, indicated by the white arrow in Figure~\ref{fig:13COC17O_pv}(b).
The PV diagram shows a spin up motion toward the center or mms1-A. 
The velocity gradient of the disk in the PV diagram  (Figure~\ref{fig:13COC17O_pv}(c)) is similar to that of the outflow (Figures~\ref{fig:pv_low} and \ref{fig:pv_intermediate}).  
Thus, we can confidentially say that the rotation of the outflow driven from mms1-A coincides with that of the disk and envelope, although we need more high-resolution data to validate the rotation profile around the disk.
Note that while \ce{C^17O}($J$=\,3\,--\,2) molecular line has some hyperfine structures, these components are degenerate because the velocity resolution of $0.9\kms$ is not high enough.
However,  the data can still provide the overall velocity structure to trace rotation motion around mms1-A.

\section{Discussion}
\label{sec:dis}

\subsection{Outflow launching radius and angular momentum}
\label{subsec:r_l}

\begin{figure}
    \centering
    \includegraphics[width=\columnwidth]{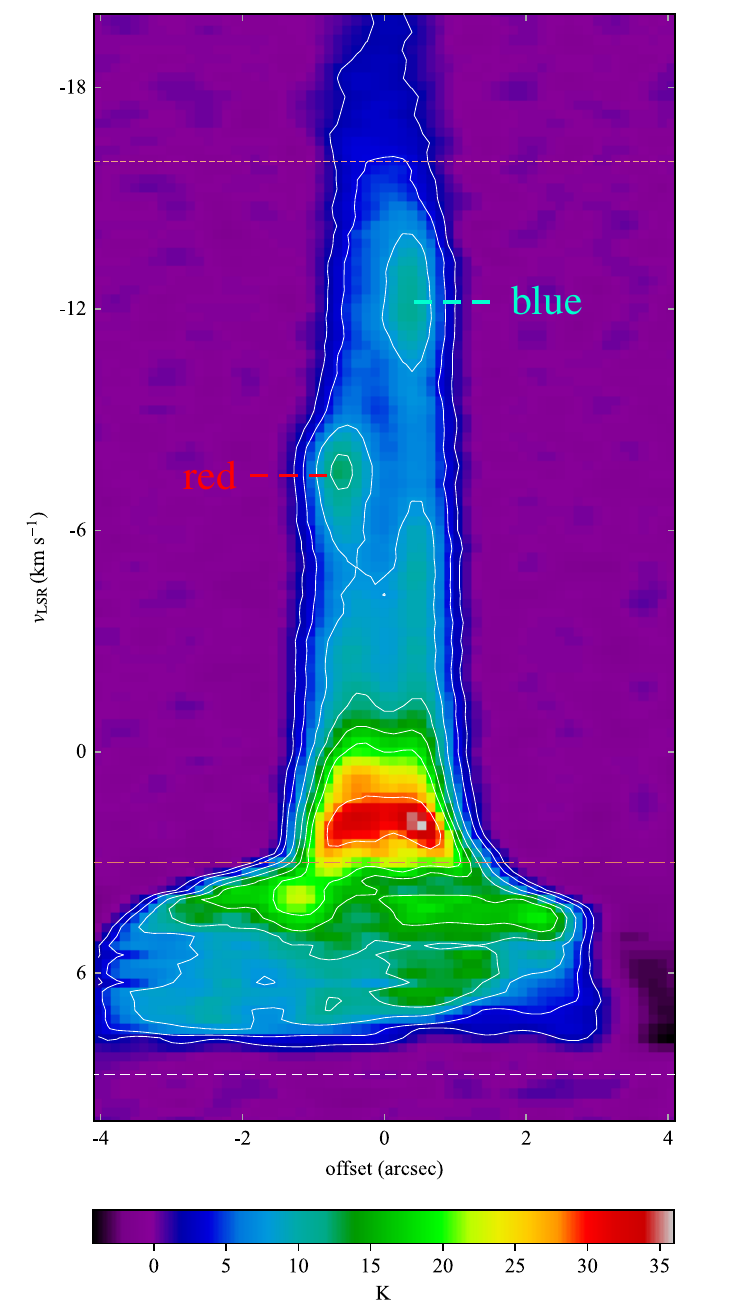}
    \caption{PV diagram on the knot B4 (see Figure~\ref{fig:knots}(c)), zooming in on the low- and intermediate-velocity components.
    The contour and horizontal lines are the same as in Figure~\ref{fig:pvmaps}.}
    \label{fig:blue_rotation}
\end{figure}

As described in \S\ref{res:rot}, we found the rotation of outflow from mms1-A.
Despite the limited spatial resolution, we could detect the outflow rotation for the low- and intermediate-velocity outflows, but we could not detect the rotation for the jets.
It is well known that protostellar outflows consist of multiple velocity components, as described in \S\ref{sec:intro}.
While almost all previous studies have detected the outflow rotation in a single velocity component, we could detect rotation in both low- and intermeviate-velocity components.
Thus, it is crucial to estimate the physical properties of the launching regions for both velocity components in order to better understand the properties of the launching object, circumstellar disk.

With the velocity gradient observed in the outflow, it is possible to estimate the physical quantities of the circumstellar disk and outflow based on Bernoulli's theorem and the conservation law of angular momentum.
According to \citet{Anderson2003ApJ}, we analytically estimated the launching radius of the outflow using observed quantities of the outflow poloidal $v_{p,\infty}$ and toroidal $v_{\phi,\infty}$ velocities and the width of the outflow $r_{\infty}$ as follows:
\begin{equation}
    \begin{split}
    r_{0} = 0.44 &\biggl( \frac{r_{\infty}}{10 \au} \biggr) ^{2/3} \biggl(\frac{v_{\phi,\infty}}{10\kms}\biggr)^{2/3}\\
    \times &\biggl(\frac{v_{p,\infty}}{100\kms}\biggr)^{-4/3}\biggl(\frac{M_{\ast}}{0.25\Msun}\biggr)^{1/3} \au.  
    \end{split}
    \label{eq:r0_rot}
\end{equation}
We applied the same method as in \citet{Matsushita2021ApJ} to measure $v_{p,\infty}$ and $v_{\phi,\infty}$ using the PV diagrams (Figures~\ref{fig:pv_low} and \ref{fig:pv_intermediate}) as  
\begin{align}
    v_{p,\infty} = \frac{1}{\cos{i}}\frac{v_\textrm{blue}+v_\textrm{red}}{2},\\
    v_{\phi,\infty} = \frac{1}{\sin{i}}\frac{v_\textrm{blue}-v_\textrm{red}}{2}.
    \label{eq:v_polphi}
\end{align}
We adopted a protostellar mass of $M_{\ast} = 0.25 \Msun$ based on the assumption in \citet{Tobin2020apj}, which is consistent with the \ce{C^17O} PV diagram (Figure~\ref{fig:13COC17O_pv}(c)), although any observations could not directly determine a dynamical mass of mms1-A.

The launching radii derived from the velocity gradient in the outflow are in the range of $67.1$--$77.1\au$ for the low-velocity outflow and $13.3$--$20.8\au$ for the intermediate-velocity outflow, as described in Table~\ref{tab:rot}, respectively. 
The launching radii of the low-velocity outflow estimated for mms1-A are larger than those for other low-mass protostars reported in previous studies \citep[e.g.,][]{Devalon2020aa,LopezVazquez2023ApJ}.
Since the radius of circumstellar disk estimated from the continuum emission is $90\pm10\au$ \citep{Tobin2020apj}, the launching radius of the low-velocity outflow is reasonably smaller thant the radius of the continuum emission.
Therefore, our observation indicates that the low-velocity outflow is driven by the outer region of the disk, as supported by both theoretical \citep{Pudritz2007prpl} and observational studies \citep[e.g.,][]{Bjerkeli2016Nature}.
As already shown in previous studies \citep[e.g.,][]{Bjerkeli2016Nature,LopezVazquez2023ApJ}, the driving radii located far from the protostar cannot be explasined by the X-wind or entrainment scenarios, which assume that a single high-velocity component of the outflow (or jet) entrains the envelope gas and produces the low-velocity components (see also \S\ref{sec:intro}).
On the other hand, the disk wind scenario provides a reasonable explanation for our observations.

According to \citet{Anderson2003ApJ}, we can estimate various physical quantities of the outflow such as the specific angular momentum and Alfv\'en radius.  
The specific angular momentum of the low-velocity outflow is in the range of $j= (0.927 {\rm -} 1.07) \times 10^{21}\cmcms$, while that of the intermediate-velocity outflow is $j= (1.27 {\rm -} 2.07) \times 10^{21}\cmcms$ (Table~\ref{tab:rot}).
It is noteworthy that the blue-shifted intermediate-velocity component seen in Figure~\ref{fig:blue_rotation} could be caused by the rotation of the outflow.
We roughly estimated its specific angular momentum from the local peaks indicated by red and blue in Figure~\ref{fig:blue_rotation}. 
With the deprojected rotational velocity $v_\phi = 3.08\kms$ and the distance of the outflow axis $193\au$, the specific angular momentum of the blue-shifted intermediate-velocity component can be estimated to be $\sim8.8 \times10^{20}\cmcms$, which is comparable to the red-shifted one. 
Note, however, that the measurement error should be significant because the blue-shifted components are more complicated than the red-shifted components.

The specific angular momentum of the outflow estimated in this study is approximately the same as that of outflows in other objects \citep[e.g.,][]{Louvet2018aa,Oya2021ApJ}. 
Previous observations have reported that prestellar cores have specific angular momenta on the order of $\sim 10^{20-21}  \cmcms \sim 100\au\kms$.
The specific angular momentum of the outflow estimated in this study is approximately the same as that of the prestellar cores \citep[e.g.,][]{Gaudel2020aa}.
This indicates that a non-negligible amount of angular momentum is cast away from the core into the interstellar space by the protostellar outflow driven from the circumstellar disk \citep[e.g.,][]{Pudritz2007prpl,Machida2013}.

The Alfv\'en radius $r_A$ and the magnetic lever arm, the ratio of the Alfv\'en radius to the launching radius $\lambda \equiv r_A/r_0$, are also described in Table~\ref{tab:rot}.
The Alfv'en radius is discribed as follows:
\begin{equation}
    r_A = \sqrt{\frac{r_\infty v_\phi}{\Omega_0}}.
    \label{eq:alfvenr}
\end{equation}
The ratio $\lambda$ for the low-velocity outflow ranges from $\lambda = 2.18$ to $2.42$, which is consistent with theoretical predictions.
On the other hand, the ratio $\lambda$ for the intermediate-velocity outflow ranges from $\lambda = 3.95$ to $4.60$, which is somewhat larger than theoretical predictions \citep{Pudritzray2019FrASS}.
The ejected material can accelerate within the Alfv\'en radius, which is determined by the ratio of magnetic pressure to ram pressure.
Therefore, it is expected that the magnetic field (or ram pressure) at the launching point is stronger (or weaker) for the intermediate-velocity outflow than for the low-velocity outflow.
This may suggest that the physical conditions at the outflow launching point and the efficiency of angular momentum transport differ between the low- and intermediate-velocity outflows.
Although further detailed observations are necessary, the observation of outflow rotation allows us to infer the physical conditions of the disk at different radii.

We adopted the protostellar mass (or dynamical mass) of HH270mms1-A as $0.25\Msun$ in the above, while Figure~\ref{fig:13COC17O_pv}(c) implies that the upper limit of the protostellar mass is about $1.0\Msun$.
However, the estimated launching radii are not significantly affected by the change in the protostellar mass, because the launching radii are proportional to the one-third power of the protostellar mass (Eq.~\ref{eq:r0_rot}). 
Assuming the protostellar mass of $1.0\Msun$, the launching radii of the low-velocity components are in the range of $106-122\au$, which exceeds the disk size estimated from the continuum emission. 
Thus, we need to determine the protostellar mass in the future in order to more accurately estimate the launching radii. 

\subsection{Circumstellar disk and high-velocity component}
\label{subsec:jet}
As described in \S\ref{res:12CO}, we could identify several knots in the jets, suggesting episodic mass ejections from the circumstellar disk.
Note that these knots may not have been spatially resolved in previous \ce{H_2} observations \citep[Figure~2 in][]{Garnavich1997ApJ}.
The knots are aligned at roughly regular intervals (Figure~\ref{fig:knots}), suggesting that their formation is not random, but depends on the activity of the mass ejections around a protostar.
Therefore, the recent history of mass ejections can be traced back from these knots.

The slight difference in the time interval between the redshifted and blueshifted components could be attributed to various factors such as measurement errors, spatial resolution limitations, or differences of each flow in inclination angles \citep[e.g.,][]{Matsushita2019ApJ}.

We consider two possible scenarios to explain the origin of the periodicity: binary interaction and disk gravitational instability. 
For the first scenario, the gravitational interactions between the binary system have a crucial role to affect the accretion onto the both protostars.
Therefore, we should estimate the orbital period of the HH270mms1 system and compare it with the time intervals of knots $\Delta t_{\rm dyn}$.
\citet{Tobin2020apj} reported that the projected separation of mms1-A and mms1-B is as large as $887\au$. 
Based on the fact that the dust mass of a circumstellar disk is proportional to the stellar mass to the power of $1.3-1.9$ \citep{Pascucci2016ApJ}, and assuming that the disk-stellar mass scaling relation is approximately $M_{\rm dust} \propto {M_{\ast}}^{1.5}$, then we can estimate the stellar mass of mms1-B and the orbital period of the HH270mms1 binary system with the stellar mass of mms1-A $0.25\Msun$.
The dust mass of mms1-A and mms1-B is $161\,M_\oplus$ and $56.7\,M_\oplus$, respectively \citep{Tobin2020apj}.
Therefore, the stellar mass of mms1-B is estimated to be $0.125\Msun$ and the orbital period to be $\sim 4.3 \times 10^4\yr$.
The time interval ($\Delta t_{\rm dyn} \sim 42 - 48\yr$) is too short to explain the periodicity due to the binary orbital motion.  
Thus, we can safely reject the binary orbital motion scenario as the origin of the periodicity.  

Next, we discuss the second scenario, gravitational instability in the circumstellar disk.
\citet{Tobin2020apj} conducted ALMA $0.87\,{\rm mm}$ and VLA $9\,{\rm mm}$ observations to measure the dust mass around mms1-A and they estimated the Toomre Q parameter \citep{Toomre1964ApJ}. 
Both observations suggest that the disk of mms1-A is massive and gravitationally unstable with $Q \leq 1 \sim 2$.
Many theoretical studies have been shown that episodic mass ejection can occur as a result of time-variable mass accretion in massive and gravitationally unstable disk \citep{Machida2014ApJ}.
In recent observations, the relation between episodic mass ejection and massive circumstellar disk has been clarified in both low-mass \citep{Lee2020NatAs} and high-mass \citep{Motogi2019ApJ} protostellar systems.
Thus, it is natural to consider that the observed knots are associated with the gravitationally unstable disk and several episodic accretions. 

The launching radius of the jet associated with the episodic accretion is expected to be smaller than those of the low-, intermediate-velocity outflows.
Due to the insufficient spatial resolution, a clear rotating signature or velocity gradient could not be observed in the jets, as described in \S~\ref{res:rot}. 
Therefore, the launching radius of the jets cannot be estimated by using the same method as for the low- and intermediate-velocity outflows.
However, using the velocity of each knot in the jets, we can estimate their launching radii based on \citet{KudohShibata1997ApJ}.
As described in \S\ref{sec:results}, we could not identify the acceleration of the jets in contrast to the lower velocity components.
The difference in the velocity structure between the jets and the lower velocity components could be attributed to the configuration of the magnetic field, in particular the ratio of the poloidal $B_{\rm p}$ to the toroidal $B_{\phi}$ components within the flow. 
\citet{KudohShibata1997ApJ} showed that the jets have a well collimated structure (so-called the magnetic tower) when the toroidal field much dominates the poloidal field ($B_{\phi}\gg B_{\rm p}$).
Note that a good collimation of the tower flow is realized due to the hoop stress \citep[see also][]{Machida2008ApJ}.
\citet{KudohShibata1997ApJ} pointed out that the velocity of such collimated or magnetic tower flows can finally reach the Keplerian velocity at their launching radius, in which the magnetic pressure gradient force as well as the magnetocentrifugal force contributes to the jet driving. 
After the reaching the terminal velocity, the jet velocity is represented by the  Keplerian velocity at the launching radius. 
Thus, the plateau velocity (or the terminal velocity) in the PV diagram (Figure~\ref{fig:pv_parallel}(a)) should approximately correspond to the Keplarian velocity at the launching radius of the knots.
The Keplerian radius $r_K$ can be estimated with the observed terminal velocity as 
\begin{equation}
    \begin{split}
    r_K &\simeq \frac{GM}{{v_{\rm K}}^{2}}\\
    &= 0.089~\biggl(\frac{M_\ast}{0.25\Msun}\biggr) \biggl( \frac{v_{\rm K}}{50 \kms}\biggr)^{-2}\ \rm{au},
    \end{split}
    \label{eq:r0_kepler}
\end{equation}
where the Keplerian velocity is replaced with the terminal velocity $v_{\rm K}=v_{\rm term}$. 
Note that the change in the terminal velocity by a factor of several does not significantly change the launching radius because the launching radius is proportional to the square root of the terminal velocity.
Thus, using equation (\ref{eq:r0_kepler}) and the velocity $v_{\rm K} = v_{\rm term}= v_\textrm{LSR}/\cos{i}$ in Tables~\ref{tab:redknots} and \ref{tab:blueknots}, the launching radii of the high-velocity components can be estimated to be $(2.36- 3.14) \times 10^{-2}\au$. 
These radii would correspond to the innermost region of the circumstellar disk \citep{Hartmann2016ARAandA}.
At such radii, it is possible that the high velocity flow or jet appears as a stellar wind from the surface of the protostar, depending on the environment at the inner region of the disk and the protostellar surface.
In such a case, contrary to our analysis, the velocity can reach several hundred $\kms$ \citep{Edwards2003ApJ}.
In addition, the ejection cycle of the stellar wind is only a few days \citep{Bouvier2020A&A}, which is much shorter than the time interval of the jet ejection described in Tables \ref{tab:redknots} and \ref{tab:blueknots}.
For these reasons, it is reasonable to consider that the jets are  driven from the inner region of the gravitationally unstable disk.

Figure~\ref{fig:pv_parallel} shows that the jets have a velocity width of $\sim 20\kms$. 
It is expected that the velocity width is comparable to or smaller than that caused by the internal bow shock \citep{Tafalla2017aa} or sideway motion \citep{Jhan2021ApJ}.
According to \citet{KudohShibata1997ApJ}, when the poloidal velocity reaches the terminal velocity, the rotational velocity of the jets is expected to be 10\% of the Keplerian velocity or less, which is too small to explain the wide velocity width of the jets.
Therefore, we can presumably rule out the origin of the velocity width as rotation.
However, if a strong magnetic fields and the long magnetic lever arm ($\lambda \sim 10$, \citealt{Matsushita2021ApJ}) are configured at the inner edge of disk associated the jets, the wide velocity width can be confirmed as rotating jets.
Future observations combined with other molecular lines such as \ce{SiO} or \ce{H_2CO}, which can trace the shock, will help to explain this velocity width \citep{Tychoniec2019aa}.

Note that the VANDAM survey in Orion \citep{Tobin2020apj} has not made the correction of $9\,{\rm mm}$ flux contamination affected by free--free emission.
We cannot estimate accurately the contribution of free--free emission to $9\,{\rm mm}$ flux because of the weak correlation between the radio luminosity and outflow force \citep{Tychoniec2018ApJS}, but the existence of EHV flows has the potential to create free--free emission near the protostar.
Thus, future centimeter observations can change the disk mass and Toomre Q parameter several times with the accurate fitting of the radio spectral energy distribution.

\subsection{Future identification of nested outflow and its substructure}
\label{subsec:identify}

\begin{figure}[htb]
    \centering
    \includegraphics[width=\columnwidth]{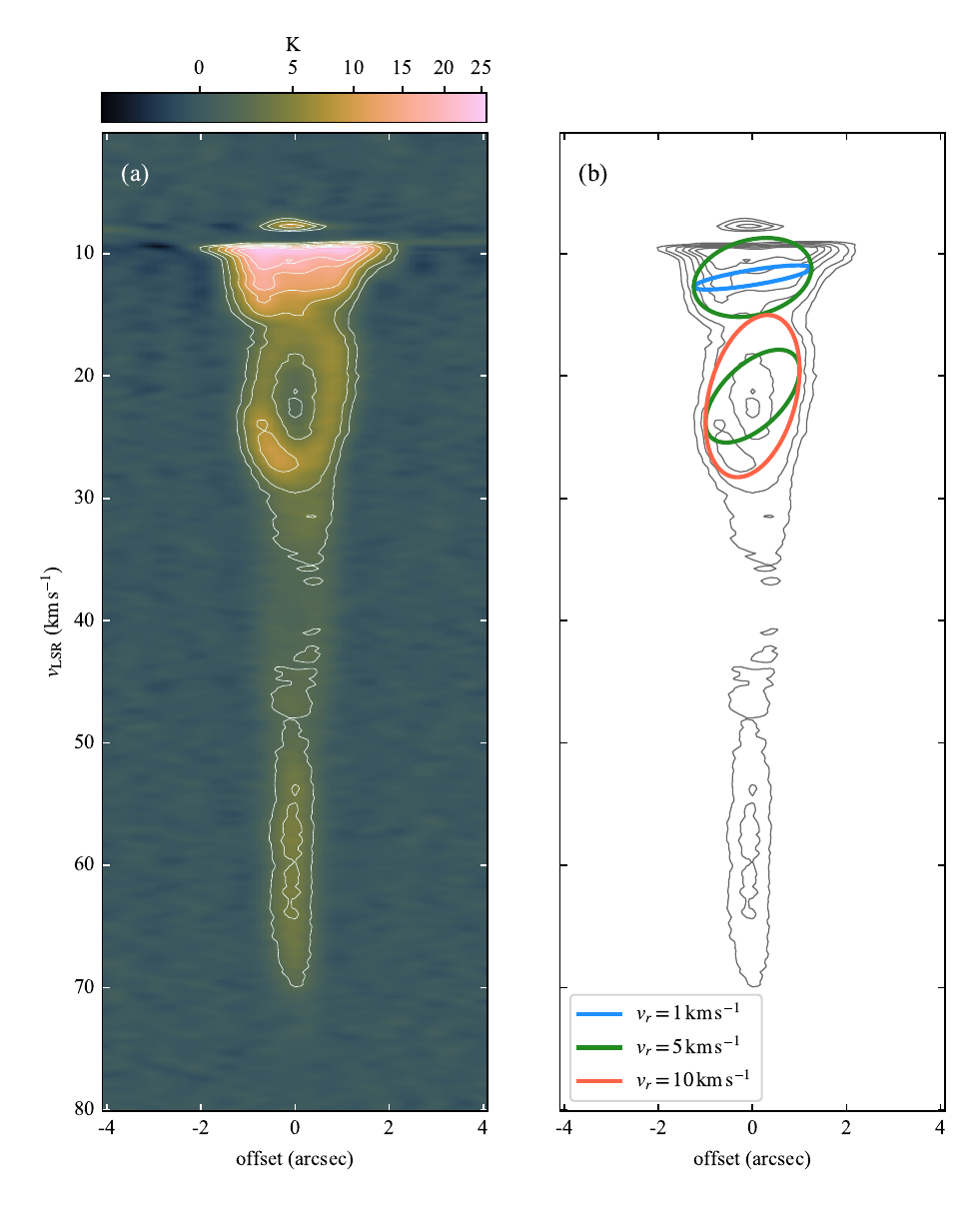}
    \caption{
    Comparison of the observation with a simple outflow model. 
    (a) PV diagram at 1200\,au far from mms1-A as in Figure~\ref{fig:pvmaps} (The difference is the maximum value of the brightness temperature).
    (b) Axisymmetric rotating outflow model constructed based on the observed quantities (Table~\ref{tab:rot}) overlapping on the contour of the PV diagram.
    The radial velocity $v_r$ is a parameter ($v_r = 1,5,10\kms$).
    For both panels, the contours indicate the levels of the brightness temperature with $[10,20,40,60,80,100,150,200] \sigma, 1\sigma = 0.19{\rm K}$.
    }
    \label{fig:ellipse}
\end{figure}

As described above, the launching region of the three-layered (i.e., low-, intermediate-, and high-velocity) flows extends from $\sim 3.0\times 10^{-2}\au$ to $77\au$. 
In addition, the foot points of the flows are not continuous but discontinuous.
Recent observations have revealed the presence of EHV flows enclosed by low-velocity wide-angle outflow \citep[e.g.,][]{Hirano2010ApJ,Matsushita2019ApJ} and layered low-velocity outflow which have different opening angles with different velocities \citep{Devalon2020aa}.
The relationship between the outflow opening angle and the outflow velocity has also been confirmed in the face-on outflow, DK Cha \citep{Harada2023ApJ}.
The outflow structure shown in the horizontal velocity gradient model \citep[see Figure~5 in][]{Harada2023ApJ} is consistent with the launching radius variation and the large-scale outflow configuration around mms1-A.
The low-velocity outflow accelerates with distance from mms1-A as reported in \citet{Lee2000ApJ} and confirmed in Figure~\ref{fig:pv_parallel}.
However, the acceleration is not significant for discussing the spatially nested velocity structure. 

The physical quantities for the outflow (outflow mass loss rate, outflow force and kinetic luminosity) around mms1-A, which are described in \S\ref{sec:outflowquantities}, are in the range of those of typical Class 0/I sources \citep{Hsieh2023ApJ}.
It should be noted that, while the acceleration was confirmed in the low- and intermediate-velocity components, we only used a typical outflow velocity to derive the physical quantities of each flow component, which differs from the approach taken by \citet{Hsieh2023ApJ}.
The error of the mass loss rate, in particular, caused by a fixed (or typical) velocity should be within several times \citep[e.g.,][]{Matsushita2018MNRAS}.
However, as shown in Table~\ref{tab:outflowquantities}, the total outflow force and kinetic luminosity are dominated by high-velocity components of which the velocities are almost constant. 
Thus, our estimates would not differ by a large order of magnitude.
As a consequence, the outflow from mms1-A is more prototypical than any outflow reported in previous observations and has the potential to explain the structures reproduced in the numerical simulations \citep[e.g.,][]{Machida2019ApJ}.

To newly identify the multi-layered structures of protostellar outflow, it is important to pay attention to the inclination angle of the circumstellar disk.
The morphological features can be easily identified in nearly edge-on system and outflows.
On the other hand, these structures are degenerate in Position--Position--Velocity (PPV) space, and thus we will not be able to identify the velocity structures of each component with certainty.
The inclination is also important in order to investigate the rotation of outflow.
As a matter of practice, it is almost impossible to detect the velocity structures indicating the rotation motion from face-on outflows.
Therefore, a moderate inclination of outflows is essential to identify both of rotating signatures and distinct velocity components, and mms1-A can be observed at an angle that can achieve this.

Figure~\ref{fig:ellipse} shows the PV diagram at $1200\au$ on which the ellipses modeled based on the outflow physical quantities in Tabel~\ref{tab:rot} are superimposed assuming the axisymmetric outflow \citep{Louvet2018aa,deValon2022aa}.
In the model, the radial velocity $v_r$ is parameterized as $v_r=1, 5, 10\kms$ and thus forms  the distorted shape (or inclined ellipse) in the PV diagram.
The ellipses indicated by the solid line roughly trace the observations where each ellipse does not overlap.
Thus, it is possible to distinguish each velocity structure (or each rotation motion of each velocity component).
As a result, we could model a similar structures to the inclined protostellar system, HH270mms1-A.

To find the nested outflow, we may need to focus on the circumstellar disk.
Recent studies have suggested that the nested outflows may be related to the substructure of the circumstellar disk, in particular the ring structure \citep[e.g.,][]{Bae2023ASPC}.
The outflow associated with a radially localized magnetic field peak can facilitate the formation of ring structures \citep{Suriano2018MNRAS,Riols2020A&A}. 
For HH270mms1, we could not detect any substructure in the continuum emission with $\sim 40\au$ resolution \citep[see also][]{Sheehan2020ApJ}.
However, the difference in the magnetic field strength \citep{Riols2020A&A} can cause the difference in the length of the magnetic lever arm $\lambda$ as described in \S\ref{subsec:r_l}.
In this study, we showed variations in the magnetic lever arm among flows, which could produce different properties of the ring structure. 
Thus, we may observe the disk with different sized rings, which are produced by magnetic effect and outflow, in future high-resolution observations.
Furthermore, similar outflows may be observed in protostellar systems where rings and gaps are already known to exist,
for instance, HL Tau \citep{ALMA2015ApJ}.

In addition to studying smaller-scale structures, focusing on larger-scale structures is also crucial for new findings.
The protostellar cores hosting the previously reported nested outflows are isolated (HH47 IRS; \citealt{Arce2013ApJ}, Barnard 5 IRS 1; \citealt{Zapata2014MNRAS} and DG Tau B \citealt{Devalon2020aa}).
The host cloud of the HH270mms1 system is also isolated. 
However, the deflected Herbig--Haro object HH270/110 around the HH270mms1 system seems to collide and interact with the outer envelope region \citep{Choi2001ApJ}.
The envelope around the protostar gradually dissipates due to the outflow, while it also affects the shape of the outflow depending on the density distribution. 
Thus, it is crucial to investigate whether the outflows can maintain their morphological characteristics through interaction with the envelope during the mass ejection stage in order to comprehend the properties of the outflows.
\citep{Arce2006ApJ,Machida2013,Offner2014ApJ}.
Therefore, the nested outflow structures would tend to be found in isolated star-forming regions rather than in the clustered regions.

\section{Conclusion}
\label{sec:conclusion}

We analyzed bipolar outflow structures associated with a protostellar binary system HH270mms1 using \ce{^12CO}($J$=\,3\,--\,2) and the isotope species \ce{^13CO}, \ce{C^17O} observed by ALMA 12m array.
Our results are summarized as follows.

\begin{enumerate}
    \item 
    The High-resolution CO and continuum observations could identify the driving source of the jets (EHV flows) and large-scale outflows as HH270mms1-A, which has a massive compact disk.
    We found several knots within the jets in the \ce{^12CO} emissions.
    The disk is sufficiently unstable to undergo time-variable mass accretion due to gravitational instability, inducing the episodic mass ejection and the formation of the chain of knots along the outflow axis, as seen in MHD numerical simulations.
    
    \item Based on morphological characteristics and velocity coherence in the PV diagram, we could identify four distinct components of low-, intermediate- and high-velocity flows and the entrained gas in the outflow associated with HH270mms1-A.
    The high- and intermediate-velocity flows are surrounded by the low-velocity flows. 
    It is considered that the outflow investigated in this study is observationally rare because similar features have been reported toward only a few objects, but a prototypical system expected from recent theoretical studies.

    \item 
    We detected rotation signatures in the low- and intermediate-velocity outflows.  
    We also confirmed that the rotation direction of the outflow is the same as in that of the envelope and disk observed in the \ce{^13CO} and \ce{C^17O} emission.
    Using an analytical method and observational properties of the outflows, we estimated the outflow launching radii.
    The launching radii of the low-velocity outflow are in the range of $67.1-77.1$\,au, while those of the intermediate-velocity outflow are in the range of $13.3-20.8$\,au. 
    This result supports the idea that these flows are driven directly from the disk by the disk wind mechanism.
    
    \item 
    Based on the observed velocity of knots and the assumption that each knot reaches its terminal velocity, corresponding to the Keplerian velocity at its driving radius in the circumstellar disk, we estimated the launching radii of the jets.
    The launching radii of the jets are distributed around the inner region of the disk, $\sim3\times10^{-2}\au$. 
    Thus, the launching radii of the jets are much smaller than those of the low- and intermediate-velocity outflows.
    
\end{enumerate}

\begin{acknowledgments}
The authors would like to appreciate the anonymous referee for constructive comments that lead to improve this work.
This paper makes use of the following ALMA data: ADS/JAO.ALMA\#2019.1.00086.S, \#2015.1.00041.S. ALMA is a partnership of ESO (representing its member states), NSF (USA) and NINS (Japan), together with NRC (Canada), MOST and ASIAA (Taiwan), and KASI (Republic of Korea), in cooperation with the Republic of Chile. The Joint ALMA Observatory is operated by ESO, AUI/NRAO and NAOJ.
This work was supported by a NAOJ ALMA Scientific Research grant Nos. 2022-22B, JSPS KAKENHI Grant Number JP21H00046, JP21K03617.
M.O. was supported by the ALMA Japan Research Grant of NAOJ ALMA Project, NAOJ-ALMA-309.
\end{acknowledgments}

\vspace{5mm}
\software{CASA \citep{2022PASP..134k4501C}, astropy \citep{2022ApJ...935..167A}, Matplotlib \citep{Huntermatplotlib}}

\appendix
\restartappendixnumbering

\section{CO channel maps and PV diagram for blueshifted outflow}
\label{apsec:figures}

\begin{figure*}[htb]
   \centering
   \includegraphics[width=\textwidth]{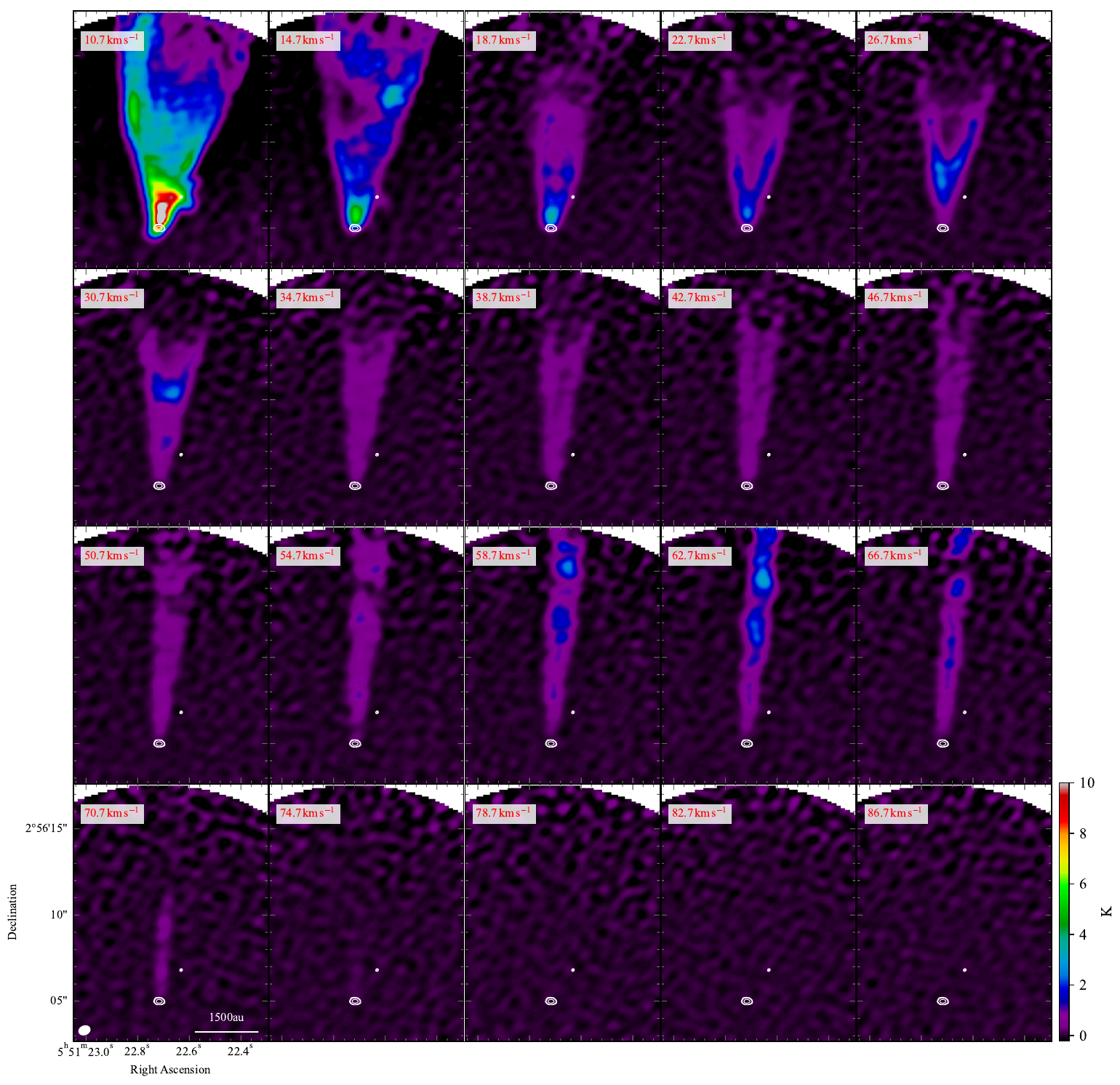}
   \caption{
   Channel maps of the \ce{^12CO} brightness temperature of the redshifted outflow from $v_\textrm{LSR} = 13.0\kms$ to $69.0\kms$ with $4.0\kms$ intervals.
   The white contours shows the continuum emission as in  Figure~\ref{fig:knots}.
   The white ellipse in the lower left corner in the bottom left panel ($53.0\kms$) indicates the synthesized beam size (Table~\ref{tab:imaging}).
   }
   \label{apfig:channelmapR_0727}
\end{figure*}

\begin{figure*}[htb]
   \centering
   \includegraphics[width=\textwidth]{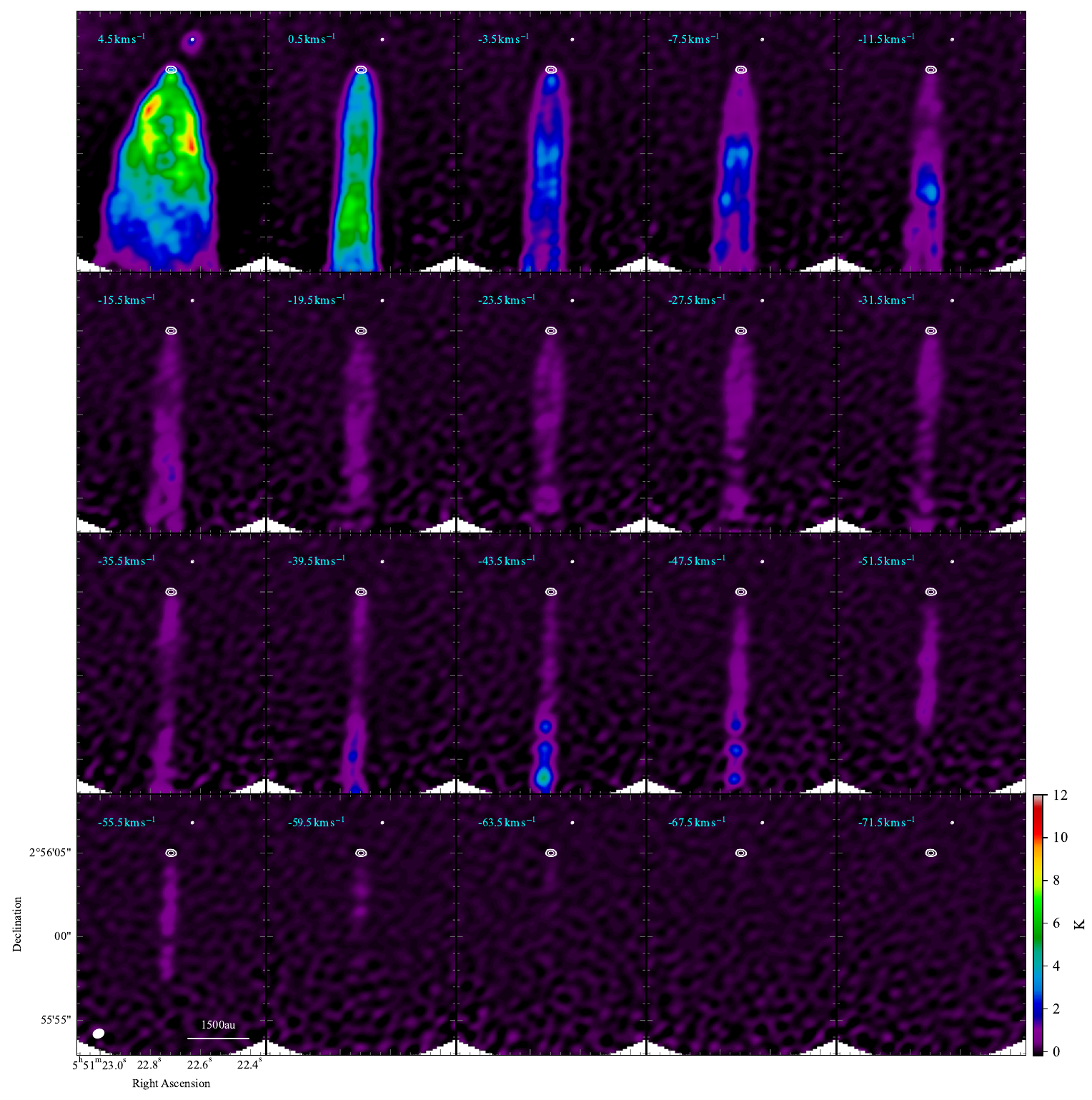}
   \caption{Same as Figure~\ref{apfig:channelmapR_0727}, but for the blueshifted outflow from $4.5\kms$ to $-71.5\kms$ with $4\kms$ intervals.}
   \label{apfig:channelmapB_0727}
\end{figure*}

\begin{figure*}[htb]
   \centering
   \includegraphics[width=\textwidth]{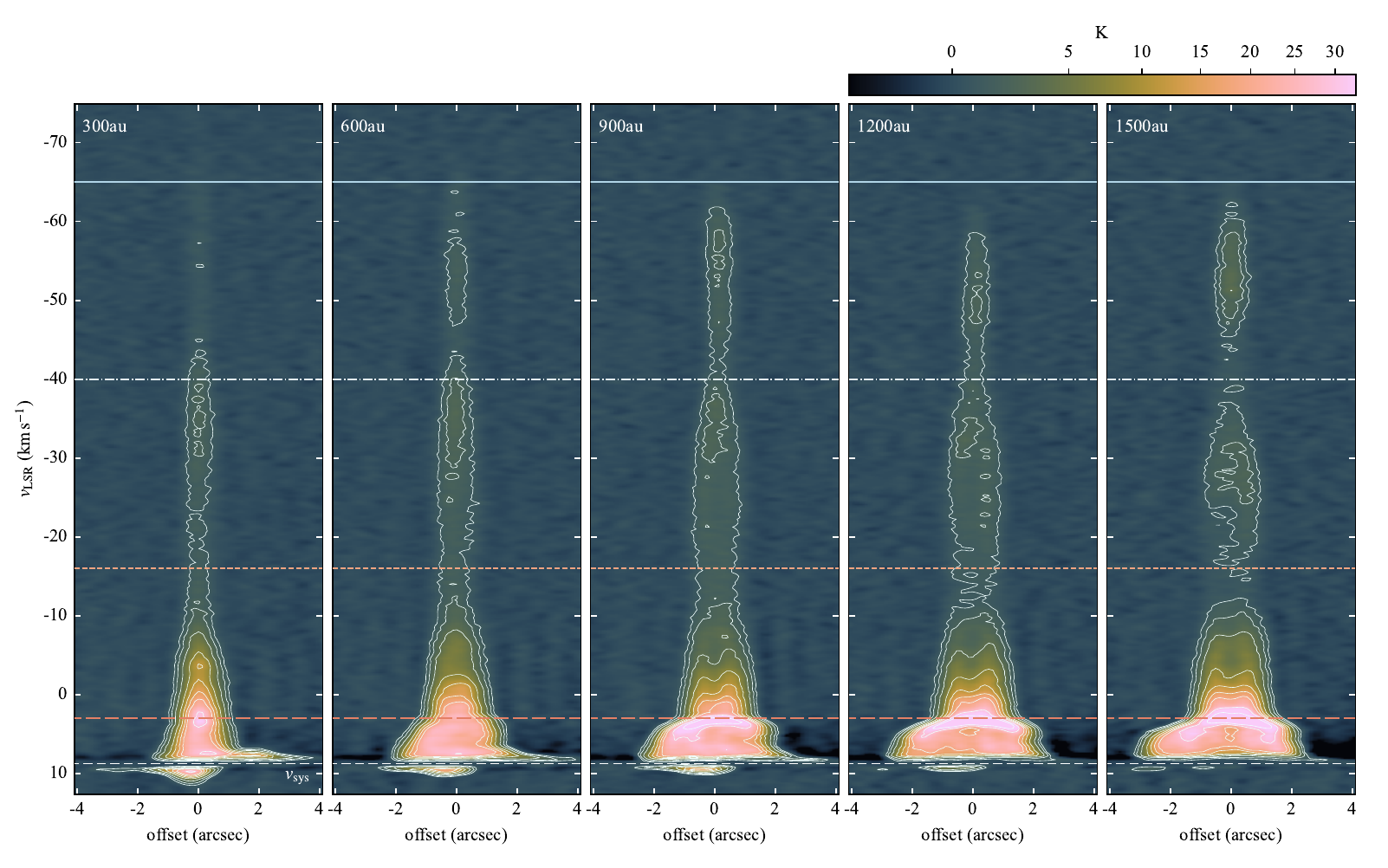}
   \caption{
   PV diagram same as Figure~\ref{fig:pvmaps} but for the blueshifted outflow.
   The contour levels of each panel are $[5,10,20,40,60,80,100,150,200]\sigma, 1\sigma = 0.19\Kdeg$.
   }
   \label{apfig:pv_blue}
\end{figure*}

We present \ce{^12CO} channel maps in Figure~\ref{apfig:channelmapR_0727} and \ref{apfig:channelmapB_0727} to confirm that the outflow components are detected over a wide velocity range.
We also show the PV diagrams perpendicular to the blue-shifted outflow in Figure~\ref{apfig:pv_blue}.
The PV diagrams of the  blueshifted outflow do not show a structure similar to the redshifted outflow such as a ring-like structure (Figure~\ref{fig:pvmaps}).

\bibliography{HH270mms1_ref}{}

\begin{thebibliography}{}
\expandafter\ifx\csname natexlab\endcsname\relax\def\natexlab#1{#1}\fi
\providecommand{\url}[1]{\href{#1}{#1}}
\providecommand{\dodoi}[1]{doi:~\href{http://doi.org/#1}{\nolinkurl{#1}}}
\providecommand{\doeprint}[1]{\href{http://ascl.net/#1}{\nolinkurl{http://ascl.net/#1}}}
\providecommand{\doarXiv}[1]{\href{https://arxiv.org/abs/#1}{\nolinkurl{https://arxiv.org/abs/#1}}}

\bibitem[{{ALMA Partnership} {et~al.}(2015){ALMA Partnership}, {Brogan},
  {P{\'e}rez}, {Hunter}, {Dent}, {Hales}, {Hills}, {Corder}, {Fomalont},
  {Vlahakis}, {Asaki}, {Barkats}, {Hirota}, {Hodge}, {Impellizzeri}, {Kneissl},
  {Liuzzo}, {Lucas}, {Marcelino}, {Matsushita}, {Nakanishi}, {Phillips},
  {Richards}, {Toledo}, {Aladro}, {Broguiere}, {Cortes}, {Cortes}, {Espada},
  {Galarza}, {Garcia-Appadoo}, {Guzman-Ramirez}, {Humphreys}, {Jung}, {Kameno},
  {Laing}, {Leon}, {Marconi}, {Mignano}, {Nikolic}, {Nyman}, {Radiszcz},
  {Remijan}, {Rod{\'o}n}, {Sawada}, {Takahashi}, {Tilanus}, {Vila Vilaro},
  {Watson}, {Wiklind}, {Akiyama}, {Chapillon}, {de Gregorio-Monsalvo}, {Di
  Francesco}, {Gueth}, {Kawamura}, {Lee}, {Nguyen Luong}, {Mangum}, {Pietu},
  {Sanhueza}, {Saigo}, {Takakuwa}, {Ubach}, {van Kempen}, {Wootten},
  {Castro-Carrizo}, {Francke}, {Gallardo}, {Garcia}, {Gonzalez}, {Hill},
  {Kaminski}, {Kurono}, {Liu}, {Lopez}, {Morales}, {Plarre}, {Schieven},
  {Testi}, {Videla}, {Villard}, {Andreani}, {Hibbard}, \&
  {Tatematsu}}]{ALMA2015ApJ}
{ALMA Partnership}, {Brogan}, C.~L., {P{\'e}rez}, L.~M., {et~al.} 2015, \apjl,
  808, L3, \dodoi{10.1088/2041-8205/808/1/L3}

\bibitem[{{Anderson} {et~al.}(2003){Anderson}, {Li}, {Krasnopolsky}, \&
  {Blandford}}]{Anderson2003ApJ}
{Anderson}, J.~M., {Li}, Z.-Y., {Krasnopolsky}, R., \& {Blandford}, R.~D. 2003,
  \apjl, 590, L107, \dodoi{10.1086/376824}

\bibitem[{{Arce} {et~al.}(2013){Arce}, {Mardones}, {Corder}, {Garay},
  {Noriega-Crespo}, \& {Raga}}]{Arce2013ApJ}
{Arce}, H.~G., {Mardones}, D., {Corder}, S.~A., {et~al.} 2013, \apj, 774, 39,
  \dodoi{10.1088/0004-637X/774/1/39}

\bibitem[{{Arce} \& {Sargent}(2006)}]{Arce2006ApJ}
{Arce}, H.~G., \& {Sargent}, A.~I. 2006, \apj, 646, 1070,
  \dodoi{10.1086/505104}

\bibitem[{{Arce} {et~al.}(2007){Arce}, {Shepherd}, {Gueth}, {Lee}, {Bachiller},
  {Rosen}, \& {Beuther}}]{Arce2007ppv}
{Arce}, H.~G., {Shepherd}, D., {Gueth}, F., {et~al.} 2007, in Protostars and
  Planets V, ed. B.~{Reipurth}, D.~{Jewitt}, \& K.~{Keil}, 245,
  \dodoi{10.48550/arXiv.astro-ph/0603071}

\bibitem[{{Astropy Collaboration} {et~al.}(2022){Astropy Collaboration},
  {Price-Whelan}, {Lim}, {Earl}, {Starkman}, {Bradley}, {Shupe}, {Patil},
  {Corrales}, {Brasseur}, {N{\"o}the}, {Donath}, {Tollerud}, {Morris},
  {Ginsburg}, {Vaher}, {Weaver}, {Tocknell}, {Jamieson}, {van Kerkwijk},
  {Robitaille}, {Merry}, {Bachetti}, {G{\"u}nther}, {Aldcroft},
  {Alvarado-Montes}, {Archibald}, {B{\'o}di}, {Bapat}, {Barentsen},
  {Baz{\'a}n}, {Biswas}, {Boquien}, {Burke}, {Cara}, {Cara}, {Conroy},
  {Conseil}, {Craig}, {Cross}, {Cruz}, {D'Eugenio}, {Dencheva}, {Devillepoix},
  {Dietrich}, {Eigenbrot}, {Erben}, {Ferreira}, {Foreman-Mackey}, {Fox},
  {Freij}, {Garg}, {Geda}, {Glattly}, {Gondhalekar}, {Gordon}, {Grant},
  {Greenfield}, {Groener}, {Guest}, {Gurovich}, {Handberg}, {Hart},
  {Hatfield-Dodds}, {Homeier}, {Hosseinzadeh}, {Jenness}, {Jones}, {Joseph},
  {Kalmbach}, {Karamehmetoglu}, {Ka{\l}uszy{\'n}ski}, {Kelley}, {Kern},
  {Kerzendorf}, {Koch}, {Kulumani}, {Lee}, {Ly}, {Ma}, {MacBride}, {Maljaars},
  {Muna}, {Murphy}, {Norman}, {O'Steen}, {Oman}, {Pacifici}, {Pascual},
  {Pascual-Granado}, {Patil}, {Perren}, {Pickering}, {Rastogi}, {Roulston},
  {Ryan}, {Rykoff}, {Sabater}, {Sakurikar}, {Salgado}, {Sanghi}, {Saunders},
  {Savchenko}, {Schwardt}, {Seifert-Eckert}, {Shih}, {Jain}, {Shukla}, {Sick},
  {Simpson}, {Singanamalla}, {Singer}, {Singhal}, {Sinha}, {Sip{\H{o}}cz},
  {Spitler}, {Stansby}, {Streicher}, {{\v{S}}umak}, {Swinbank}, {Taranu},
  {Tewary}, {Tremblay}, {Val-Borro}, {Van Kooten}, {Vasovi{\'c}}, {Verma}, {de
  Miranda Cardoso}, {Williams}, {Wilson}, {Winkel}, {Wood-Vasey}, {Xue},
  {Yoachim}, {Zhang}, {Zonca}, \& {Astropy Project
  Contributors}}]{2022ApJ...935..167A}
{Astropy Collaboration}, {Price-Whelan}, A.~M., {Lim}, P.~L., {et~al.} 2022,
  \apj, 935, 167, \dodoi{10.3847/1538-4357/ac7c74}

\bibitem[{{Bacciotti} {et~al.}(2002){Bacciotti}, {Ray}, {Mundt},
  {Eisl{\"o}ffel}, \& {Solf}}]{Bacciotti2002ApJ}
{Bacciotti}, F., {Ray}, T.~P., {Mundt}, R., {Eisl{\"o}ffel}, J., \& {Solf}, J.
  2002, \apj, 576, 222, \dodoi{10.1086/341725}

\bibitem[{{Bae} {et~al.}(2023){Bae}, {Isella}, {Zhu}, {Martin}, {Okuzumi}, \&
  {Suriano}}]{Bae2023ASPC}
{Bae}, J., {Isella}, A., {Zhu}, Z., {et~al.} 2023, in Astronomical Society of
  the Pacific Conference Series, Vol. 534, Astronomical Society of the Pacific
  Conference Series, ed. S.~{Inutsuka}, Y.~{Aikawa}, T.~{Muto}, K.~{Tomida}, \&
  M.~{Tamura}, 423

\bibitem[{{Banerjee} \& {Pudritz}(2006)}]{BanerjeePudritz2006ApJ}
{Banerjee}, R., \& {Pudritz}, R.~E. 2006, \apj, 641, 949,
  \dodoi{10.1086/500496}

\bibitem[{{Bjerkeli} {et~al.}(2016){Bjerkeli}, {van der Wiel}, {Harsono},
  {Ramsey}, \& {J{\o}rgensen}}]{Bjerkeli2016Nature}
{Bjerkeli}, P., {van der Wiel}, M. H.~D., {Harsono}, D., {Ramsey}, J.~P., \&
  {J{\o}rgensen}, J.~K. 2016, \nat, 540, 406, \dodoi{10.1038/nature20600}

\bibitem[{{Blandford} \& {Payne}(1982)}]{Blandfordpayne1982MNRAS}
{Blandford}, R.~D., \& {Payne}, D.~G. 1982, \mnras, 199, 883,
  \dodoi{10.1093/mnras/199.4.883}

\bibitem[{{Bouvier} {et~al.}(2020){Bouvier}, {Alecian}, {Alencar}, {Sousa},
  {Donati}, {Perraut}, {Bayo}, {Rebull}, {Dougados}, {Duvert}, {Berger},
  {Benisty}, {Pouilly}, {Folsom}, {Moutou}, \& {SPIRou
  Consortium}}]{Bouvier2020A&A}
{Bouvier}, J., {Alecian}, E., {Alencar}, S.~H.~P., {et~al.} 2020, \aap, 643,
  A99, \dodoi{10.1051/0004-6361/202038892}

\bibitem[{{CASA Team} {et~al.}(2022){CASA Team}, {Bean}, {Bhatnagar}, {Castro},
  {Donovan Meyer}, {Emonts}, {Garcia}, {Garwood}, {Golap}, {Gonzalez Villalba},
  {Harris}, {Hayashi}, {Hoskins}, {Hsieh}, {Jagannathan}, {Kawasaki},
  {Keimpema}, {Kettenis}, {Lopez}, {Marvil}, {Masters}, {McNichols},
  {Mehringer}, {Miel}, {Moellenbrock}, {Montesino}, {Nakazato}, {Ott}, {Petry},
  {Pokorny}, {Raba}, {Rau}, {Schiebel}, {Schweighart}, {Sekhar}, {Shimada},
  {Small}, {Steeb}, {Sugimoto}, {Suoranta}, {Tsutsumi}, {van Bemmel},
  {Verkouter}, {Wells}, {Xiong}, {Szomoru}, {Griffith}, {Glendenning}, \&
  {Kern}}]{2022PASP..134k4501C}
{CASA Team}, {Bean}, B., {Bhatnagar}, S., {et~al.} 2022, \pasp, 134, 114501,
  \dodoi{10.1088/1538-3873/ac9642}

\bibitem[{{Choi}(2001)}]{Choi2001ApJ}
{Choi}, M. 2001, \apj, 550, 817, \dodoi{10.1086/319777}

\bibitem[{{Davis} {et~al.}(1994){Davis}, {Mundt}, \&
  {Eisloeffel}}]{Davis1994ApJ}
{Davis}, C.~J., {Mundt}, R., \& {Eisloeffel}, J. 1994, \apjl, 437, L55,
  \dodoi{10.1086/187681}

\bibitem[{{de Valon} {et~al.}(2020){de Valon}, {Dougados}, {Cabrit}, {Louvet},
  {Zapata}, \& {Mardones}}]{Devalon2020aa}
{de Valon}, A., {Dougados}, C., {Cabrit}, S., {et~al.} 2020, \aap, 634, L12,
  \dodoi{10.1051/0004-6361/201936950}

\bibitem[{{de Valon} {et~al.}(2022){de Valon}, {Dougados}, {Cabrit}, {Louvet},
  {Zapata}, \& {Mardones}}]{deValon2022aa}
---. 2022, \aap, 668, A78, \dodoi{10.1051/0004-6361/202141316}

\bibitem[{{Dunham} {et~al.}(2014){Dunham}, {Arce}, {Mardones}, {Lee},
  {Matthews}, {Stutz}, \& {Williams}}]{Dunham2014ApJ}
{Dunham}, M.~M., {Arce}, H.~G., {Mardones}, D., {et~al.} 2014, \apj, 783, 29,
  \dodoi{10.1088/0004-637X/783/1/29}

\bibitem[{{Edwards} {et~al.}(2003){Edwards}, {Fischer}, {Kwan}, {Hillenbrand},
  \& {Dupree}}]{Edwards2003ApJ}
{Edwards}, S., {Fischer}, W., {Kwan}, J., {Hillenbrand}, L., \& {Dupree}, A.~K.
  2003, \apjl, 599, L41, \dodoi{10.1086/381077}

\bibitem[{{Feddersen} {et~al.}(2020){Feddersen}, {Arce}, {Kong}, {Suri},
  {S{\'a}nchez-Monge}, {Ossenkopf-Okada}, {Dunham}, {Nakamura}, {Shimajiri}, \&
  {Bally}}]{Feddersen2020ApJ}
{Feddersen}, J.~R., {Arce}, H.~G., {Kong}, S., {et~al.} 2020, \apj, 896, 11,
  \dodoi{10.3847/1538-4357/ab86a9}

\bibitem[{{Frerking} {et~al.}(1982){Frerking}, {Langer}, \&
  {Wilson}}]{Frerking1982ApJ}
{Frerking}, M.~A., {Langer}, W.~D., \& {Wilson}, R.~W. 1982, \apj, 262, 590,
  \dodoi{10.1086/160451}

\bibitem[{{Garnavich} {et~al.}(1997){Garnavich}, {Noriega-Crespo}, {Raga}, \&
  {B{\"o}hm}}]{Garnavich1997ApJ}
{Garnavich}, P.~M., {Noriega-Crespo}, A., {Raga}, A.~C., \& {B{\"o}hm}, K.-H.
  1997, \apj, 490, 752, \dodoi{10.1086/304887}

\bibitem[{{Gaudel} {et~al.}(2020){Gaudel}, {Maury}, {Belloche}, {Maret},
  {Andr{\'e}}, {Hennebelle}, {Galametz}, {Testi}, {Cabrit}, {Palmeirim},
  {Ladjelate}, {Codella}, \& {Podio}}]{Gaudel2020aa}
{Gaudel}, M., {Maury}, A.~J., {Belloche}, A., {et~al.} 2020, \aap, 637, A92,
  \dodoi{10.1051/0004-6361/201936364}

\bibitem[{{Ginsburg} {et~al.}(2011){Ginsburg}, {Bally}, \&
  {Williams}}]{Ginsburg2011MNRAS}
{Ginsburg}, A., {Bally}, J., \& {Williams}, J.~P. 2011, \mnras, 418, 2121,
  \dodoi{10.1111/j.1365-2966.2011.19279.x}

\bibitem[{{Harada} {et~al.}(2023){Harada}, {Tokuda}, {Yamasaki}, {Sato},
  {Omura}, {Hirano}, {Onishi}, {Tachihara}, \& {Machida}}]{Harada2023ApJ}
{Harada}, N., {Tokuda}, K., {Yamasaki}, H., {et~al.} 2023, \apj, 945, 63,
  \dodoi{10.3847/1538-4357/acb930}

\bibitem[{{Hartmann} {et~al.}(2016){Hartmann}, {Herczeg}, \&
  {Calvet}}]{Hartmann2016ARAandA}
{Hartmann}, L., {Herczeg}, G., \& {Calvet}, N. 2016, \araa, 54, 135,
  \dodoi{10.1146/annurev-astro-081915-023347}

\bibitem[{{Hirano} {et~al.}(2010){Hirano}, {Ho}, {Liu}, {Shang}, {Lee}, \&
  {Bourke}}]{Hirano2010ApJ}
{Hirano}, N., {Ho}, P. P.~T., {Liu}, S.-Y., {et~al.} 2010, \apj, 717, 58,
  \dodoi{10.1088/0004-637X/717/1/58}

\bibitem[{{Hirota} {et~al.}(2017){Hirota}, {Machida}, {Matsushita}, {Motogi},
  {Matsumoto}, {Kim}, {Burns}, \& {Honma}}]{Hirota2017NatAs}
{Hirota}, T., {Machida}, M.~N., {Matsushita}, Y., {et~al.} 2017, Nature
  Astronomy, 1, 0146, \dodoi{10.1038/s41550-017-0146}

\bibitem[{{Hsieh} {et~al.}(2023){Hsieh}, {Arce}, {Li}, {Dunham}, {Offner},
  {Stephens}, {Stutz}, {Megeath}, {Kong}, {Plunkett}, {Tobin}, {Zhang},
  {Mardones}, {Pineda}, {Stanke}, \& {Carpenter}}]{Hsieh2023ApJ}
{Hsieh}, C.-H., {Arce}, H.~G., {Li}, Z.-Y., {et~al.} 2023, \apj, 947, 25,
  \dodoi{10.3847/1538-4357/acba13}

\bibitem[{{Hull} {et~al.}(2016){Hull}, {Girart}, {Kristensen}, {Dunham},
  {Rodr{\'\i}guez-Kamenetzky}, {Carrasco-Gonz{\'a}lez}, {Cort{\'e}s}, {Li}, \&
  {Plambeck}}]{Hull2016ApJ}
{Hull}, C. L.~H., {Girart}, J.~M., {Kristensen}, L.~E., {et~al.} 2016, \apjl,
  823, L27, \dodoi{10.3847/2041-8205/823/2/L27}

\bibitem[{Hunter(2007)}]{Huntermatplotlib}
Hunter, J.~D. 2007, Computing in Science \& Engineering, 9, 90,
  \dodoi{10.1109/MCSE.2007.55}

\bibitem[{{Jhan} \& {Lee}(2021)}]{Jhan2021ApJ}
{Jhan}, K.-S., \& {Lee}, C.-F. 2021, \apj, 909, 11,
  \dodoi{10.3847/1538-4357/abd6c5}

\bibitem[{{Kounkel} {et~al.}(2017){Kounkel}, {Hartmann}, {Loinard},
  {Ortiz-Le{\'o}n}, {Mioduszewski}, {Rodr{\'\i}guez}, {Dzib}, {Torres}, {Pech},
  {Galli}, {Rivera}, {Boden}, {Evans}, {Brice{\~n}o}, \&
  {Tobin}}]{Kounkel2017ApJ}
{Kounkel}, M., {Hartmann}, L., {Loinard}, L., {et~al.} 2017, \apj, 834, 142,
  \dodoi{10.3847/1538-4357/834/2/142}

\bibitem[{{Kounkel} {et~al.}(2018){Kounkel}, {Covey}, {Su{\'a}rez},
  {Rom{\'a}n-Z{\'u}{\~n}iga}, {Hernandez}, {Stassun}, {Jaehnig}, {Feigelson},
  {Pe{\~n}a Ram{\'\i}rez}, {Roman-Lopes}, {Da Rio}, {Stringfellow}, {Kim},
  {Borissova}, {Fern{\'a}ndez-Trincado}, {Burgasser},
  {Garc{\'\i}a-Hern{\'a}ndez}, {Zamora}, {Pan}, \& {Nitschelm}}]{Kounkel2018AJ}
{Kounkel}, M., {Covey}, K., {Su{\'a}rez}, G., {et~al.} 2018, \aj, 156, 84,
  \dodoi{10.3847/1538-3881/aad1f1}

\bibitem[{{Kudoh} \& {Shibata}(1997)}]{KudohShibata1997ApJ}
{Kudoh}, T., \& {Shibata}, K. 1997, \apj, 474, 362, \dodoi{10.1086/303437}

\bibitem[{{Launhardt} {et~al.}(2009){Launhardt}, {Pavlyuchenkov}, {Gueth},
  {Chen}, {Dutrey}, {Guilloteau}, {Henning}, {Pi{\'e}tu}, {Schreyer}, \&
  {Semenov}}]{Laundhardt2009aa}
{Launhardt}, R., {Pavlyuchenkov}, Y., {Gueth}, F., {et~al.} 2009, \aap, 494,
  147, \dodoi{10.1051/0004-6361:200810835}

\bibitem[{{Lee}(2020)}]{Lee2020A&ARv}
{Lee}, C.-F. 2020, \aapr, 28, 1, \dodoi{10.1007/s00159-020-0123-7}

\bibitem[{{Lee} {et~al.}(2017){Lee}, {Ho}, {Li}, {Hirano}, {Zhang}, \&
  {Shang}}]{Lee2017NatAs}
{Lee}, C.-F., {Ho}, P. T.~P., {Li}, Z.-Y., {et~al.} 2017, Nature Astronomy, 1,
  0152, \dodoi{10.1038/s41550-017-0152}

\bibitem[{{Lee} {et~al.}(2020){Lee}, {Li}, \& {Turner}}]{Lee2020NatAs}
{Lee}, C.-F., {Li}, Z.-Y., \& {Turner}, N.~J. 2020, Nature Astronomy, 4, 142,
  \dodoi{10.1038/s41550-019-0905-x}

\bibitem[{{Lee} {et~al.}(2000){Lee}, {Mundy}, {Reipurth}, {Ostriker}, \&
  {Stone}}]{Lee2000ApJ}
{Lee}, C.-F., {Mundy}, L.~G., {Reipurth}, B., {Ostriker}, E.~C., \& {Stone},
  J.~M. 2000, \apj, 542, 925, \dodoi{10.1086/317056}

\bibitem[{{L{\'o}pez-V{\'a}zquez}
  {et~al.}(2023{\natexlab{a}}){L{\'o}pez-V{\'a}zquez}, {Lee},
  {Fern{\'a}ndez-L{\'o}pez}, {Louvet}, {Guerra-Alvarado}, \&
  {Zapata}}]{Lopes2023}
{L{\'o}pez-V{\'a}zquez}, J.~A., {Lee}, C.-F., {Fern{\'a}ndez-L{\'o}pez}, M.,
  {et~al.} 2023{\natexlab{a}}, arXiv e-prints, arXiv:2312.03272,
  \dodoi{10.48550/arXiv.2312.03272}

\bibitem[{{L{\'o}pez-V{\'a}zquez}
  {et~al.}(2023{\natexlab{b}}){L{\'o}pez-V{\'a}zquez}, {Zapata}, \&
  {Lee}}]{LopezVazquez2023ApJ}
{L{\'o}pez-V{\'a}zquez}, J.~A., {Zapata}, L.~A., \& {Lee}, C.-F.
  2023{\natexlab{b}}, \apj, 944, 63, \dodoi{10.3847/1538-4357/acb439}

\bibitem[{{Louvet} {et~al.}(2018){Louvet}, {Dougados}, {Cabrit}, {Mardones},
  {M{\'e}nard}, {Tabone}, {Pinte}, \& {Dent}}]{Louvet2018aa}
{Louvet}, F., {Dougados}, C., {Cabrit}, S., {et~al.} 2018, \aap, 618, A120,
  \dodoi{10.1051/0004-6361/201731733}

\bibitem[{{Machida}(2014)}]{Machida2014ApJ}
{Machida}, M.~N. 2014, \apjl, 796, L17, \dodoi{10.1088/2041-8205/796/1/L17}

\bibitem[{{Machida} \& {Basu}(2019)}]{Machida2019ApJ}
{Machida}, M.~N., \& {Basu}, S. 2019, \apj, 876, 149,
  \dodoi{10.3847/1538-4357/ab18a7}

\bibitem[{{Machida} \& {Hosokawa}(2013)}]{Machida2013}
{Machida}, M.~N., \& {Hosokawa}, T. 2013, \mnras, 431, 1719,
  \dodoi{10.1093/mnras/stt291}

\bibitem[{{Machida} {et~al.}(2008){Machida}, {Inutsuka}, \&
  {Matsumoto}}]{Machida2008ApJ}
{Machida}, M.~N., {Inutsuka}, S.-i., \& {Matsumoto}, T. 2008, \apj, 676, 1088,
  \dodoi{10.1086/528364}

\bibitem[{{Machida} {et~al.}(2004){Machida}, {Tomisaka}, \&
  {Matsumoto}}]{Machida2004}
{Machida}, M.~N., {Tomisaka}, K., \& {Matsumoto}, T. 2004, \mnras, 348, L1,
  \dodoi{10.1111/j.1365-2966.2004.07402.x}

\bibitem[{{Matsushita} {et~al.}(2018){Matsushita}, {Sakurai}, {Hosokawa}, \&
  {Machida}}]{Matsushita2018MNRAS}
{Matsushita}, Y., {Sakurai}, Y., {Hosokawa}, T., \& {Machida}, M.~N. 2018,
  \mnras, 475, 391, \dodoi{10.1093/mnras/stx3070}

\bibitem[{{Matsushita} {et~al.}(2021){Matsushita}, {Takahashi}, {Ishii},
  {Tomisaka}, {Ho}, {Carpenter}, \& {Machida}}]{Matsushita2021ApJ}
{Matsushita}, Y., {Takahashi}, S., {Ishii}, S., {et~al.} 2021, \apj, 916, 23,
  \dodoi{10.3847/1538-4357/ac069f}

\bibitem[{{Matsushita} {et~al.}(2019){Matsushita}, {Takahashi}, {Machida}, \&
  {Tomisaka}}]{Matsushita2019ApJ}
{Matsushita}, Y., {Takahashi}, S., {Machida}, M.~N., \& {Tomisaka}, K. 2019,
  \apj, 871, 221, \dodoi{10.3847/1538-4357/aaf1b6}

\bibitem[{{Motogi} {et~al.}(2019){Motogi}, {Hirota}, {Machida}, {Yonekura},
  {Honma}, {Takakuwa}, \& {Matsushita}}]{Motogi2019ApJ}
{Motogi}, K., {Hirota}, T., {Machida}, M.~N., {et~al.} 2019, \apjl, 877, L25,
  \dodoi{10.3847/2041-8213/ab212f}

\bibitem[{{Offner} \& {Arce}(2014)}]{Offner2014ApJ}
{Offner}, S. S.~R., \& {Arce}, H.~G. 2014, \apj, 784, 61,
  \dodoi{10.1088/0004-637X/784/1/61}

\bibitem[{{Oya} {et~al.}(2018){Oya}, {Sakai}, {Watanabe}, {L{\'o}pez-Sepulcre},
  {Ceccarelli}, {Lefloch}, \& {Yamamoto}}]{Oya2018ApJ}
{Oya}, Y., {Sakai}, N., {Watanabe}, Y., {et~al.} 2018, \apj, 863, 72,
  \dodoi{10.3847/1538-4357/aacf42}

\bibitem[{{Oya} {et~al.}(2021){Oya}, {Watanabe}, {L{\'o}pez-Sepulcre},
  {Ceccarelli}, {Lefloch}, {Favre}, \& {Yamamoto}}]{Oya2021ApJ}
{Oya}, Y., {Watanabe}, Y., {L{\'o}pez-Sepulcre}, A., {et~al.} 2021, \apj, 921,
  12, \dodoi{10.3847/1538-4357/ac0a72}

\bibitem[{{Pascucci} {et~al.}(2016){Pascucci}, {Testi}, {Herczeg}, {Long},
  {Manara}, {Hendler}, {Mulders}, {Krijt}, {Ciesla}, {Henning}, {Mohanty},
  {Drabek-Maunder}, {Apai}, {Sz{\H{u}}cs}, {Sacco}, \&
  {Olofsson}}]{Pascucci2016ApJ}
{Pascucci}, I., {Testi}, L., {Herczeg}, G.~J., {et~al.} 2016, \apj, 831, 125,
  \dodoi{10.3847/0004-637X/831/2/125}

\bibitem[{{Pelletier} \& {Pudritz}(1992)}]{PeeletierPudritz1992ApJ}
{Pelletier}, G., \& {Pudritz}, R.~E. 1992, \apj, 394, 117,
  \dodoi{10.1086/171565}

\bibitem[{{Pudritz} \& {Norman}(1986)}]{Pudritznorman1986ApJ}
{Pudritz}, R.~E., \& {Norman}, C.~A. 1986, \apj, 301, 571,
  \dodoi{10.1086/163924}

\bibitem[{{Pudritz} {et~al.}(2007){Pudritz}, {Ouyed}, {Fendt}, \&
  {Brandenburg}}]{Pudritz2007prpl}
{Pudritz}, R.~E., {Ouyed}, R., {Fendt}, C., \& {Brandenburg}, A. 2007, in
  Protostars and Planets V, ed. B.~{Reipurth}, D.~{Jewitt}, \& K.~{Keil}, 277,
  \dodoi{10.48550/arXiv.astro-ph/0603592}

\bibitem[{{Pudritz} \& {Ray}(2019)}]{Pudritzray2019FrASS}
{Pudritz}, R.~E., \& {Ray}, T.~P. 2019, Frontiers in Astronomy and Space
  Sciences, 6, 54, \dodoi{10.3389/fspas.2019.00054}

\bibitem[{{Reipurth} \& {Olberg}(1991)}]{Reipurth1991aa}
{Reipurth}, B., \& {Olberg}, M. 1991, \aap, 246, 535

\bibitem[{{Riols} {et~al.}(2020){Riols}, {Lesur}, \& {Menard}}]{Riols2020A&A}
{Riols}, A., {Lesur}, G., \& {Menard}, F. 2020, \aap, 639, A95,
  \dodoi{10.1051/0004-6361/201937418}

\bibitem[{{Rodr{\'i}guez} {et~al.}(1998){Rodr{\'i}guez}, {Reipurth}, {Raga}, \&
  {Cant{\'o}}}]{Rodriguez1998RMxAA}
{Rodr{\'i}guez}, L.~F., {Reipurth}, B., {Raga}, A.~C., \& {Cant{\'o}}, J. 1998,
  \rmxaa, 34, 69

\bibitem[{{Sep{\'u}lveda} {et~al.}(2011){Sep{\'u}lveda}, {Anglada},
  {Estalella}, {L{\'o}pez}, {Girart}, \& {Yang}}]{Sepuleveda2011aa}
{Sep{\'u}lveda}, I., {Anglada}, G., {Estalella}, R., {et~al.} 2011, \aap, 527,
  A41, \dodoi{10.1051/0004-6361/200912916}

\bibitem[{{Sheehan} {et~al.}(2020){Sheehan}, {Tobin}, {Federman}, {Megeath}, \&
  {Looney}}]{Sheehan2020ApJ}
{Sheehan}, P.~D., {Tobin}, J.~J., {Federman}, S., {Megeath}, S.~T., \&
  {Looney}, L.~W. 2020, \apj, 902, 141, \dodoi{10.3847/1538-4357/abbad5}

\bibitem[{{Shu} {et~al.}(1994){Shu}, {Najita}, {Ostriker}, {Wilkin}, {Ruden},
  \& {Lizano}}]{Shu1994ApJ}
{Shu}, F., {Najita}, J., {Ostriker}, E., {et~al.} 1994, \apj, 429, 781,
  \dodoi{10.1086/174363}

\bibitem[{{Snell} {et~al.}(1980){Snell}, {Loren}, \& {Plambeck}}]{Snell1980ApJ}
{Snell}, R.~L., {Loren}, R.~B., \& {Plambeck}, R.~L. 1980, \apjl, 239, L17,
  \dodoi{10.1086/183283}

\bibitem[{{Stone} \& {Norman}(1993)}]{Stonenorman1993ApJ}
{Stone}, J.~M., \& {Norman}, M.~L. 1993, \apj, 413, 210, \dodoi{10.1086/172989}

\bibitem[{{Suriano} {et~al.}(2018){Suriano}, {Li}, {Krasnopolsky}, \&
  {Shang}}]{Suriano2018MNRAS}
{Suriano}, S.~S., {Li}, Z.-Y., {Krasnopolsky}, R., \& {Shang}, H. 2018, \mnras,
  477, 1239, \dodoi{10.1093/mnras/sty717}

\bibitem[{{Tabone} {et~al.}(2017){Tabone}, {Cabrit}, {Bianchi}, {Ferreira},
  {Pineau des For{\^e}ts}, {Codella}, {Gusdorf}, {Gueth}, {Podio}, \&
  {Chapillon}}]{Tabone2017aa}
{Tabone}, B., {Cabrit}, S., {Bianchi}, E., {et~al.} 2017, \aap, 607, L6,
  \dodoi{10.1051/0004-6361/201731691}

\bibitem[{{Tafalla} {et~al.}(2017){Tafalla}, {Su}, {Shang}, {Johnstone},
  {Zhang}, {Santiago-Garc{\'\i}a}, {Lee}, {Hirano}, \& {Wang}}]{Tafalla2017aa}
{Tafalla}, M., {Su}, Y.~N., {Shang}, H., {et~al.} 2017, \aap, 597, A119,
  \dodoi{10.1051/0004-6361/201629493}

\bibitem[{{Takahashi} {et~al.}(2008){Takahashi}, {Saito}, {Ohashi}, {Kusakabe},
  {Takakuwa}, {Shimajiri}, {Tamura}, \& {Kawabe}}]{Takahashi2008ApJ}
{Takahashi}, S., {Saito}, M., {Ohashi}, N., {et~al.} 2008, \apj, 688, 344,
  \dodoi{10.1086/592212}

\bibitem[{{Tobin} {et~al.}(2020){Tobin}, {Sheehan}, {Megeath},
  {D{\'\i}az-Rodr{\'\i}guez}, {Offner}, {Murillo}, {van 't Hoff}, {van
  Dishoeck}, {Osorio}, {Anglada}, {Furlan}, {Stutz}, {Reynolds}, {Karnath},
  {Fischer}, {Persson}, {Looney}, {Li}, {Stephens}, {Chandler}, {Cox},
  {Dunham}, {Tychoniec}, {Kama}, {Kratter}, {Kounkel}, {Mazur}, {Maud},
  {Patel}, {Perez}, {Sadavoy}, {Segura-Cox}, {Sharma}, {Stephenson}, {Watson},
  \& {Wyrowski}}]{Tobin2020apj}
{Tobin}, J.~J., {Sheehan}, P.~D., {Megeath}, S.~T., {et~al.} 2020, \apj, 890,
  130, \dodoi{10.3847/1538-4357/ab6f64}

\bibitem[{{Tokuda} {et~al.}(2022){Tokuda}, {Zahorecz}, {Kunitoshi},
  {Higashino}, {Tanaka}, {Konishi}, {Suzuki}, {Kitano}, {Harada}, {Shimonishi},
  {Neelamkodan}, {Fukui}, {Kawamura}, {Onishi}, \& {Machida}}]{Tokuda2022ApJL}
{Tokuda}, K., {Zahorecz}, S., {Kunitoshi}, Y., {et~al.} 2022, \apjl, 936, L6,
  \dodoi{10.3847/2041-8213/ac81c1}

\bibitem[{{Tomida} {et~al.}(2010){Tomida}, {Tomisaka}, {Matsumoto}, {Ohsuga},
  {Machida}, \& {Saigo}}]{Tomida2010ApJ}
{Tomida}, K., {Tomisaka}, K., {Matsumoto}, T., {et~al.} 2010, \apjl, 714, L58,
  \dodoi{10.1088/2041-8205/714/1/L58}

\bibitem[{{Tomisaka}(2000)}]{Tomisaka2000ApJ}
{Tomisaka}, K. 2000, \apjl, 528, L41, \dodoi{10.1086/312417}

\bibitem[{{Toomre}(1964)}]{Toomre1964ApJ}
{Toomre}, A. 1964, \apj, 139, 1217, \dodoi{10.1086/147861}

\bibitem[{{Tsukamoto} {et~al.}(2022){Tsukamoto}, {Maury}, {Commer{\c{c}}on},
  {Alves}, {Cox}, {Sakai}, {Ray}, {Zhao}, \& {Machida}}]{Tsukamoto2022ppvii}
{Tsukamoto}, Y., {Maury}, A., {Commer{\c{c}}on}, B., {et~al.} 2022, arXiv
  e-prints, arXiv:2209.13765, \dodoi{10.48550/arXiv.2209.13765}

\bibitem[{{Tychoniec} {et~al.}(2019){Tychoniec}, {Hull}, {Kristensen}, {Tobin},
  {Le Gouellec}, \& {van Dishoeck}}]{Tychoniec2019aa}
{Tychoniec}, {\L}., {Hull}, C. L.~H., {Kristensen}, L.~E., {et~al.} 2019, \aap,
  632, A101, \dodoi{10.1051/0004-6361/201935409}

\bibitem[{{Tychoniec} {et~al.}(2018){Tychoniec}, {Tobin}, {Karska}, {Chandler},
  {Dunham}, {Harris}, {Kratter}, {Li}, {Looney}, {Melis}, {P{\'e}rez},
  {Sadavoy}, {Segura-Cox}, \& {van Dishoeck}}]{Tychoniec2018ApJS}
{Tychoniec}, {\L}., {Tobin}, J.~J., {Karska}, A., {et~al.} 2018, \apjs, 238,
  19, \dodoi{10.3847/1538-4365/aaceae}

\bibitem[{{Wang} {et~al.}(2019){Wang}, {Shang}, \& {Chiang}}]{Wang2019ApJ}
{Wang}, L.-Y., {Shang}, H., \& {Chiang}, T.-Y. 2019, \apj, 874, 31,
  \dodoi{10.3847/1538-4357/ab07b5}

\bibitem[{{Zapata} {et~al.}(2014){Zapata}, {Arce}, {Brassfield}, {Palau},
  {Patel}, \& {Pineda}}]{Zapata2014MNRAS}
{Zapata}, L.~A., {Arce}, H.~G., {Brassfield}, E., {et~al.} 2014, \mnras, 441,
  3696, \dodoi{10.1093/mnras/stu810}

\bibitem[{{Zapata} {et~al.}(2015){Zapata}, {Lizano}, {Rodr{\'\i}guez}, {Ho},
  {Loinard}, {Fern{\'a}ndez-L{\'o}pez}, \& {Tafoya}}]{Zapata2015ApJ}
{Zapata}, L.~A., {Lizano}, S., {Rodr{\'\i}guez}, L.~F., {et~al.} 2015, \apj,
  798, 131, \dodoi{10.1088/0004-637X/798/2/131}

\bibitem[{{Zhang} {et~al.}(2018){Zhang}, {Higuchi}, {Sakai}, {Oya},
  {L{\'o}pez-Sepulcre}, {Imai}, {Sakai}, {Watanabe}, {Ceccarelli}, {Lefloch},
  \& {Yamamoto}}]{Zhang2018ApJ}
{Zhang}, Y., {Higuchi}, A.~E., {Sakai}, N., {et~al.} 2018, \apj, 864, 76,
  \dodoi{10.3847/1538-4357/aad7ba}

\end{thebibliography}
\bibliographystyle{aasjournal}
\end{document}